# The effect of turbulence on the minimum thickness of a liquid film flowing down a vertical tube

David T. Hughes [†]

Formerly at School of Chemical Engineering,
University of Birmingham, Edgbaston,
Birmingham B15 2TT, U.K.

Abstract

When a liquid flows down a vertical tube at a low flowrate, it may not be able to completely cover the surface as a film, and the flow will comprise a number of rivulets separated by unwetted areas. Predictions of the minimum film thickness and corresponding minimum flowrate below which films cannot be sustained are needed in many industrial heat and mass transfer operations.  In this paper a model is developed using the minimum energy method to calculate the minimum film thicknesses that can be maintained for flows down a vertical tube. It is concluded that inclusion of the effect of turbulence in the model can significantly increase the values of the calculated minimum film thicknesses.  In these calculations (for 100ºC water flowing down a 6mm inside diameter tube) the effect of turbulence increased as the tube to liquid contact angle increased. For a flow at the highest contact angle of 90º, the minimum film thickness increased by 22%  above the laminar value and the corresponding minimum flowrate increased by 75 %. The results of the calculations are compared to experimental data obtained from a falling film evaporator.

Nomenclature

| | |
|---|---|
| $A$ | area |
| $C_k, C_{\omega 1}, C_{\omega 2}$ | constants in equations (8) and (9) |
| $C_p$ | specific heat capacity |
| $e$ | element number |
| $E_{K1}$ | kinetic energy of the mean flow per unit length of tube |
| $E_{K2}$ | kinetic energy of the turbulence per unit length of tube |

[†] Email: davidhughes5005@gmail.com

| | |
|---|---|
| $E_T$ | total energy per unit length of tube |
| $F$ | volumetric flowrate |
| $F_T$ | total volumetric flowrate |
| (film) | for a complete film |
| $g$ | acceleration due to gravity |
| $h_{ext}$ | external heat transfer coefficient |
| $k$ | specific turbulence kinetic energy |
| $k_i, k_j, k_k$ | specific turbulence kinetic energy at nodes $i, j, k$ |
| $K$ | von Karman constant |
| $Ka$ | Kapitza number |
| $m$ | total number of elements |
| $n$ | number of rivulets |
| $n_0$ | outer normal of the free surface of the flow |
| $N_i, N_j, N_k$ | shape functions for nodes $i, j, k$ |
| $P_k$ | production of turbulence kinetic energy |
| $P_L$ | pressure in the liquid phase |
| $P_V$ | pressure in the vapour phase |
| $Pr, Pr_t$ | Prandtl number, turbulent Prandtl number |
| $Q$ | heat transfer rate per unit length of tube |
| $R$ | internal radius of tube |
| $Re, Re_{crit}$ | Reynolds number, critical Reynolds number (equation (11)) |
| (riv) | for a single rivulet |
| $S$ | radius of curvature of rivulet |
| $T_0$ | temperature of the liquid free surface |
| $u_z$ | time averaged fluid velocity down tube |
| $u_{z_i}, u_{z_j}, u_{z_k}$ | time averaged fluid velocity at nodes $i, j, k$ |
| $u^*$ | friction velocity |
| $U$ | overall heat transfer coefficient |
| $w$ | width |
| $x$ | cartesian coordinate |
| $X$ | wetted to total tube area |
| $y$ | cartesian coordinate |
| $y_p$ | perpendicular distance from wall |
| $y^+$ | dimensionless distance from wall |
| $z$ | cartesian coordinate |

Greek symbols

| | |
|---|---|
| $\alpha$ | half angle subtended by rivulet at tube axis |
| $\beta$ | angle (Fig 1) |
| $\gamma_{SL}, \gamma_{SV}, \gamma_{LV}$ | surface tension, solid-liquid, solid-vapour, liquid-vapour |
| $\Gamma$ | mass flowrate per unit width |
| $\delta_f$ | film thickness |
| $\delta_{max}$ | rivulet centreline thickness (Fig 1) |
| $\delta_p$ | perpendicular distance of liquid free surface from wall |
| $\delta_s$ | tube wall thickness |
| $\delta_{turb}$ | film thickness at which turbulence first appears |
| $\delta^+$ | dimensionless distance of free surface from wall |
| $\delta_*$ | critical film thickness |
| $\Delta_*$ | dimensionless critical film thickness |
| $\Delta T$ | temperature difference |
| $\theta$ | contact angle |
| $\lambda$ | thermal conductivity of fluid |
| $\lambda'$ | thermal conductivity |
| $\lambda_s$ | thermal conductivity of solid tube wall |
| $\Lambda$ | $(\rho^3 g^2 / 15 \mu^2 \gamma_{LV})^{1/5}$ |
| $\mu$ | dynamic viscosity |
| $\nu$ | kinematic viscosity |
| $\nu_t$ | kinematic turbulent viscosity |
| $\rho$ | fluid density |
| $\sigma_k, \sigma_\omega,$ | constants in equations (8) and (9) |
| $\tau_w$ | wall shear stress |
| $\Phi$ | function (Appendix C) |
| $\omega$ | specific dissipation rate |

Subscripts

| | |
|---|---|
| * | critical |

Superscripts

| | |
|---|---|
| (e) | element |

Abbreviations

| | |
|---|---|
| FB | Force Balance |
| MEP | Mudawar and El-Masri profile (equation 11) |
| MTE | Minimum Total Energy |

1. Introduction

When a liquid flows down a vertical solid surface at a low flowrate, it may not be able to completely cover the surface as a film, and the flow will comprise a number of rivulets separated by unwetted areas. Predictions of the minimum film thickness ($\delta_*$) and minimum flowrate ($F_*$) below which it is not possible to sustain a falling film are needed in many industrial heat and mass transfer operations in the chemical, petroleum refining, desalination, and food industries. Plant optimisation can sometimes depend critically on the optimum operation of the falling film equipment. Operation of falling film equipment at a flowrate significantly above or below $F_*$ may increase running costs or reduce plant performance.

The breakdown of falling films can be modelled using the force balance method and the minimum energy method. Hartley and Murgatroyd [1] obtained an expression for $\delta_*$ using a force balance at the upstream stagnation point of a dry patch. This model can be used for laminar and turbulent flows. If the film is laminar the minimum film thickness is found to be related to the liquid-solid wall contact angle (*θ)* by the equation

$$\delta_* = \frac{(1 - cos\theta)^{1/5}}{\Lambda} \qquad (1)$$

where $\Lambda$ is a function of the physical properties of the liquid and is given by $(\rho^3 g^2/15\mu^2 \gamma_{LV})^{1/5}$. Equation (1) shows one of the main features of both the force balance and minimum energy methods which is that an increase in $\delta_*$ is predicted as the contact angle increases.

In the minimum energy method, the total (surface plus kinetic) energies of film and rivulet flows are calculated and compared. Mikielewicz and Moszynski [4] built on the models of Hobler [2] and Bankhoff [3] and derived an expression for the minimum thickness of a film flowing down a plane wall by assuming that the break-up of the film into rivulets occurred when both the continuous film and the rivulets had the same mass flow and total energy and when the energy per unit width of the rivulet configuration had a local minimum. A summary of these models ([1], [2] and [4]) is presented in Table 1.

The minimum total energy (MTE) method has also been used to investigate the effect of shear stresses at the liquid free surface by Saber and El-Genk [5] and Wilson, Sullivan, and Duffy [6] and the effect of wall corrugations has been modelled by Junqing Wei, Jinping Liu and Xiongwen Xu [7]. Hughes and Bott [8] developed a model based on the work of Mikielewicz and Moszynski [4] and investigated the effect of wall curvature. This model can be applied to the flow of fluids down tubes and is directly relevant to many industrial processes. At very low flow rates trains of sliding drops are observed (Schmuki and Laso [9]) and the rivulet to sliding drop transition has been modelled using the minimum energy method by Ghezzehei [10].

In both the force balance and minimum energy methods the velocity fields in falling films have to be calculated and in minimum energy calculations the velocity fields in the rivulets have also to be calculated. Laminar falling films can be modelled using the Nusselt equation and for turbulent falling films, Mudawar and El-Masri [11] have developed a semi empirical profile to model the turbulent viscosity across the film. In this profile the turbulent viscosity is damped as free liquid surface is approached.

Rivulet flows can be calculated numerically and also by using the lubrication approximation. Allen and Biggin [12] used the finite element method to obtain velocity distributions in a rivulet flowing down an inclined plane. The flowrates predicted by numerical models for rivulets with a contact angle of 90º flowing down a plane vertical wall can be checked against an analytic solution which has been derived by Perazzo and Gratton [13]. The flow of a non-Newtonian rivulet down a vertical surface has been investigated by Al Mukahal, Duffy and Wilson [14]. Lopes, Borges , Matias, Pimenta and Siqueira [15] have investigated the accuracy of lubrication models for rivulets and a have developed a new lubrication model with improved accuracy and enlarged range of applicability.

In both film and rivulet flows, surface waves have been observed to occur. Mascarenhas and Mudawar[16] have investigated the influence of interfacial waves on the mass, momentum and heat transfer in turbulent, free-falling water films and the wave structure of rivulets has been studied by Alekseenko, Bobylev, Guzanov, Markovich and Kharlamov [17] over a range of Reynolds number and contact angles.

In industrial falling film evaporators stainless steel tubes are often used because stainless steel has good resistance to corrosion and excellent strength over a wide range of temperatures. Because of this, a number of experiments have been conducted to determine the minimum wetting rate of liquids, particularly water, on vertical steel surfaces. The values obtained in these experiments depend on the contact angle and can depend on the method of liquid distribution (Morison, Worth and O'dea [18] and Lu, Stehmann, Yuan and Scholl [19]). In addition, Dreiser and Bart [20] have shown that wetting hysteresis can occur in which the flowrate required to completely wet the surface, when the flowrate is being increased from a low value, is higher than the flowrate associated with film breakdown when the flowrate is being reduced from a high value (an established film being in place at the high flowrate).

Values of minimum wetting rate that have been reported in the literature for water on a vertical steel surface are presented in terms of the mass flowrate per unit width of surface ($\Gamma_*$). Paramalingam, Winchester and Marsh [21] obtained a value of 0.222 kg m$^{-1}$ s$^{-1}$ for 20 ºC water. Munakata, Watanabe and Miyashita [22] reported values from approximately 0.065 to 0.12 kg m$^{-1}$ s$^{-1}$. Morison and Tandon [23] obtained values of $\Gamma_*$ that decreased from 0.16 kg m$^{-1}$ s$^{-1}$ to 0.12 kg m$^{-1}$ s$^{-1}$ as the water temperature increased from 20 ºC to 70 ºC. Morison et al. [18] obtained a value of 0.104 kg m$^{-1}$ s$^{-1}$ for water at both 25 ºC and 60 ºC.

In studies in which the contact angle is also known, it is possible to compare the experimental results with the values predicted by the force balance and minimum energy models. Morison et al.[18] determined the minimum wetting rates of water and aqueous solutions of glycerol, alcohol and calcium chloride in a 1 m long, 48 mm internal diameter, vertical, stainless steel tube. These fluids were chosen because they have a wide range of liquid physical properties and contact angle (64º to 98°). The film thicknesses calculated from the force balance model were always greater than the minimum film thicknesses found in the experiments. Mikielewicz and Moszynski [4] compared the results of their minimum energy model with the results of Hobler [2] in which the minimum thicknesses were measured for water on a variety of surfaces (aluminium, glass, copper, stainless steel and varnish) . The contact angles were in range 35.8 to 56.8 degrees. It was found that the predictions of film thickness from theory using the minimum energy method were uniformly too low by a factor of 1.5.

In the models ([4] to [10]) in which a minimum total (surface plus kinetic) energy criterion is used, the flows are assumed to be laminar, and the total energy is taken to be the sum of a surface component and volumetric component, the volumetric component being the kinetic energy of the flow. Fedorchenko and Hruby [24] and Fedorchenko and Abdulkhalikov [25] put forward a modified expression for the volumetric component and this increased the predicted values of $\delta_*$ for flow down a plane wall. These increased values of $\delta_*$ coincided with the experimental values (of Hobler[2], presented by Mikielewicz and Moszynski [4]) with an average accuracy of 5%.

The purpose of the present paper is to investigate whether turbulence has an effect on the calculation of the minimum film thickness for flows down a vertical tube and whether there is a dependence on the liquid-tube contact angle. In this work, in which the minimum energy method is used, the volumetric component is assumed the be the kinetic energy of the flow and the effect of turbulence, if present, is to modify the velocity distributions in the flows. In previous calculations on this system performed by the author it was assumed that the rivulet and film flows were always laminar [8]. The results from the calculations in this paper are also used to model the evaporation of water at 100 °C flowing down the inside of a 6 mm tube with a water-tube contact angle of 50°. These values for the water temperature, tube diameter and contact angle have been chosen because they represent the parameters of a system for which experimental results are available [26]. The experimental programme that generated these results was initiated to study the feasibility of the use of plastic tubes in vapour recompression falling film evaporators for desalination and other duties [27- 29]. Further details of the experiment and the evaporator are presented in the Appendix.

The main focus in this paper is on the minimum total energy (MTE) method but results from the force balance (FB) method are also calculated and compared with the experimental data.

## 2. Overview of the calculation procedure

The liquid flow down the vertical tube is assumed to comprise either a continuous film or a single rivulet or several identical rivulets. At a given flowrate, the configuration the system adopts is decided by considering the total energy of all the possible configurations in which the system can exist. The configuration that the system adopts is assumed to be the configuration with the lowest energy. The lowest energy states are found by plotting all the configurations in which the system can exist on a graph of total energy versus total flowrate. Each possible configuration is represented by a point on the graph. At a given flowrate the lowest energy state can be identified from the graph. The calculation procedure has following stages:

1. The 2D steady state fully developed velocity distributions for films and rivulets travelling down the vertical tube are calculated over a range of film thickness ($\delta_f$) and range of rivulet size (using the variable $\alpha$, which is the half angle subtended by the rivulet at the tube axis). These calculations are performed for three cases: Case1 (no turbulence), Case2 (turbulence modelled using the $k$-$\omega$ model) and Case 3 (turbulence modelled using a turbulent viscosity profile). The contact angle ($\theta$) was initially set to 50º as this was the contact angle in the evaporator experiments.
2. The film and rivulet flowrates and kinetic energies are calculated from the film and rivulet velocity distributions. These results are presented in graphs of flowrate and energy as a function of the film thickness ($\delta_f$) and rivulet half angle ($\alpha$). Each graph has up to three curves, each curve being associated with one of the turbulence cases.
3. The film and rivulet energies are combined with expressions for surface energy to give values of the total energy for each flow configuration. Expressions for the total flowrate are also obtained. These data are plotted on a graph of total energy versus total flowrate. Each possible flow configuration appears as a point on this graph.
4. The lowest energy configuration at each flowrate (which is the expected configuration) is identified from the total energy-total flowrate graph by inspection. The flowrate below which it is not expected that a film can be sustained is found from this graph.
5. A total energy-total flowrate graph is constructed for a number of different values of contact angle, and the variation of minimum film thickness with contact angle is found from these graphs.
6. In order to simulate the thermal performance of the evaporator, the heat transfer characteristics of the flow configurations are calculated by incorporating the flow geometries into thermal models that simulate the conditions in the plastic tube evaporator experiments (described in the Appendix). The results of these calculations are presented in graphs of heat transfer rate as a function of the film thickness ($\delta_f$) and rivulet half angle ($\alpha$).

7. The values of $\delta_f$ or $\alpha$ (and the number of rivulets) associated with the lowest energy configurations are found at each flowrate from the data used to construct the energy flowrate graph (for $\theta$ = 50°). These data are then used with the heat transfer graphs to calculate the rate of heat transfer associated with each lowest energy configuration.

8. A graph of heat transfer coefficient versus total flowrate is constructed and this is compared with the experimental results from the plastic tube evaporator.

The volumetric flowrate of the flows is a variable that occurs frequently in this paper. The Reynolds number for the flow is given by $4\Gamma/\mu$ where $\Gamma$ is the mass flowrate per unit length of the tube circumference. Using the physical property data that are used in these calculations for this system (which are presented in the Appendix) and the tube inside diameter of 6 mm, it is calculated that a Reynolds number of 2000 corresponds to a film flowrate of 2784 mm³ s⁻¹. Reynolds numbers are used for both film and rivulet flows.

## 3 Fluid Flow Calculations

### 3.1 Calculation of the film and rivulet velocity distributions

A cartesian set of coordinates is used in which the film and rivulet flows have a time averaged velocity $u_z(x, y)$ in the direction of the $z$ axis. The $z$ axis is coincident with the tube axis and is the direction in which gravity acts. For a fully developed incompressible flow under steady state conditions, there is no time averaged flow in a plane perpendicular to the $z$ axis and the momentum equation for turbulent flow is given by

$$\frac{\partial}{\partial x}\left[(\nu + \nu_t)\frac{\partial u_z}{\partial x}\right] + \frac{\partial}{\partial y}\left[(\nu + \nu_t)\frac{\partial u_z}{\partial y}\right] + g = 0 \qquad (2)$$

where $\nu$ is the kinematic laminar viscosity and $\nu_t$ is the kinematic turbulent viscosity. The film and rivulet velocity distributions are calculated by solving equation (2) over the fluid domains using the finite element method.

### 3.1.1 Film Flow

In the film flow calculations, a section of the film of thickness $\delta_f$ is meshed as shown in Fig. 1, where $\delta_f$ is the film thickness and $R$ is the inside radius of the tube.

### 3.1.2 Rivulet flow

For a rivulet flowing down a vertical tube the radius of curvature ($S$) at each point of the rivulet surface is given by

$$P_L - P_V = \frac{\gamma_{LV}}{S} \qquad (3)$$

where $P_L$ is the pressure in the liquid phase, $P_V$ is the pressure in the vapour phase and $\gamma_{LV}$ is the surface tension. On a vertical surface, $P_L$ is a constant across the rivulet and $P_V$ is constant. The free surface of the rivulet has a constant temperature $T_0$ and so $\gamma_{LV}$ is constant. Therefore, from equation (3), the rivulet free surface has the same radius of curvature at each point on the surface. The rivulet free surface is an arc of a circle of radius $S$ as shown in Fig 1.

The rivulet contacts the tube surface at an angle $\theta$ (the contact angle) and subtends an angle $2\alpha$ at the centre of the tube of internal radius $R$. The rivulet free surface subtends an angle $2\beta$ at the centre of the circle of radius $S$ and the centreline thickness of the rivulet is $\delta_{max}$. For rivulets with the same value of contact angle $\theta$, larger rivulets have larger values of $S$. When $\theta$ is greater than $\alpha$, the rivulet has a biconvex shape as in Fig 1. When $\alpha$ is greater that $\theta$, the rivulet shape is concavo-convex. In the calculations presented in this paper, all rivulets are biconvex. From Fig. 1, it can be seen that

$$\theta = \alpha + \beta \qquad (4)$$

$$S \sin \beta = R \sin \alpha \qquad (5)$$

and

$$\delta_{max} = R(1 - \cos \alpha) + S(1 - \cos \beta) \qquad (6)$$

In the rivulet flow calculations, because of symmetry, only half the rivulet is meshed as shown in Fig. 1.

3.2 Incorporation of turbulence into the calculation

The effect of turbulence is modelled by incorporating a turbulent viscosity $v_t$ in the film and rivulet calculations. Turbulence is generated by the shearing action of the fluid and is damped at the tube wall. In a falling film, turbulence is also damped at the liquid free surface (Kharangate, Lee and Mudawar [30]). $v_t$ is zero both at the tube wall and at the liquid free surface. The effect of turbulence is investigated by considering the following three cases:

    Case 1 - no turbulence.
    Case 2 - $v_t$ is calculated using the $k$-$\omega$ model.

Case 3 - $v_t$ is calculated from a semi-empirical turbulent viscosity profile.

In Case1, which is the laminar case, $v_t$ is taken to be zero at all points.
In Case 2, the $k$-$\omega$ model is used to calculate $v_t$ from the equation

$$v_t = \frac{k}{\omega} \tag{7}$$

where $k$ and $\omega$, the specific turbulence kinetic energy and specific dissipation rate, are found from the 2D steady state equations

$$\frac{\partial}{\partial x}\left[(v + \sigma_k v_t)\frac{\partial k}{\partial x}\right] + \frac{\partial}{\partial y}\left[(v + \sigma_k v_t)\frac{\partial k}{\partial y}\right] + P_k - C_k \omega k = 0 \tag{8}$$

and

$$\frac{\partial}{\partial x}\left[(v + \sigma_\omega v_t)\frac{\partial \omega}{\partial x}\right] + \frac{\partial}{\partial y}\left[(v + \sigma_\omega v_t)\frac{\partial \omega}{\partial y}\right] + C_{\omega 1}\frac{\omega}{k} P_k - C_{\omega 2}\omega^2 = 0 \tag{9}$$

where $P_k$, the production of turbulence kinetic energy, is given by

$$P_k = v_t \left[\left(\frac{\partial u_z}{\partial x}\right)^2 + \left(\frac{\partial u_z}{\partial y}\right)^2\right] \tag{10}$$

The equation constants have the values, $C_k = 0.09$, $C_{\omega 1} = 0.52$, $C_{\omega 2} = 0.0708$, $\sigma_k = 0.6$, $\sigma_\omega = 0.5$. (Wilcox [31]). It is known that the results from $k$-$\omega$ models have a strong dependency on the free stream values of $\omega$ outside the shear layer. In order to reduce this effect, $k$-$\omega$ models can be blended with a $k$-epsilon model, as in the shear stress transport (SST) model (Menter[32]). However, the flows presented in this paper are all fully developed and so there are no free stream regions in these simulations. It is not necessary to blend the $k$-$\omega$ model with a $k$-epsilon model in these calculations.

In case 3, the Mudawar and El-Masri profile (MEP) [11], [30] is used to calculate $v_t$. This is a 1D profile which describes the variation of the turbulent viscosity across a turbulent falling film. This semi-empirical profile shows good agreement with experimental data for freely falling films undergoing heating or evaporation and has been shown to be particularly successful in the low-turbulence region [11]. The MEP equation for $v_t$ is:

$$\frac{v_t}{v} = -\frac{1}{2} + \frac{1}{2}\sqrt{1 + 4K^2(y^+)^2\left(1 - \frac{y^+}{\delta^+}\right)^2\left\{1 - exp\left[-\frac{y^+}{26}\left(1 - \frac{y^+}{\delta^+}\right)^{1/2}\left(1 - \frac{0.865 Re_{crit}^{1/2}}{\delta^+}\right)\right]\right\}^2} \tag{11}$$

where $K$ is the von Karman constant (taken to be 0.4) and $Re_{crit} = 0.04/Ka^{0.37}$ for films undergoing evaporation. The Kapitza number ($Ka$) is equal to $\mu^4 g/\rho \gamma_{LV}^3$. Using the physical property data for water at 100 °C presented in the Appendix, $Re_{crit}$ is calculated to be 1672. The dimensionless distance from the wall $y^+$ is equal to $y_p u^*/\nu$ and the dimensionless distance of the free surface from the wall $\delta^+$ is given by $\delta_p u^*/\nu$ where $y_p$ is the distance to the wall and $\delta_p$ is the distance of the liquid free surface to the wall. Also, $u^*$ is the friction velocity at the wall ($\sqrt{\tau_w/\rho}$). When $\delta^+$ is less than $0.865\, Re_{crit}^{1/2}$, the flow is laminar. $\nu_t$ is zero both at the wall and at the liquid free surface.

From a force balance on a fully developed film flowing down a vertical tube, the friction velocity ($u^*$) at the tube wall is found to be equal to $\sqrt{\delta_p\left(1 - \frac{\delta_p}{2R}\right)g}$. And so, for a film flowing down a plane wall ($R=\infty$), the friction velocity at the wall is equal to $\sqrt{g\delta_p}$. From these equations, the film thickness at which turbulence first appears $\delta_{turb}$ can be written as

$$\frac{\delta_{turb}\sqrt{\delta_{turb}\left(1 - \frac{\delta_{turb}}{2R}\right)g}}{\nu} = 0.865\, Re_{crit}^{1/2} \qquad (12)$$

For 100 °C water $\delta_{turb}$ is calculated to be 226 μm when $R$ = 3mm and 223 μm when $R = \infty$.

In this paper it is assumed that the 1D MEP equation (11) can be used to obtain viscosity profiles in the 2D films and rivulets. In order to use this profile to find the viscosity at a point P in a film or rivulet, a line is drawn through the point P from the tube wall to the tube axis; $y_p$ is the perpendicular distance from the tube wall to P, $\delta_p$ is the perpendicular distance from the tube wall to the film or rivulet free surface and $u^*$ is the value of the friction velocity at the point where the line contacts the tube wall. With these values of values of $y_p$, $\delta_p$ and $u^*$, the values of $y^+$ and $\delta^+$ are calculated. These values of $y^+$ and $\delta^+$ are then substituted into equation (11) to give a value of $\nu_t$ at point P. This approximation is similar to the approximation made in model of Mikielewicz and Moszynski [4] in which the rivulet velocity distribution is calculated by dividing the rivulet into thin strips and then assuming that the velocity profile in each strip is the same as the velocity distribution in a uniform film of the same thickness. The method by which the wall friction velocities along the tube wall were calculated is described in the next section.

3.3 Solution of the flow equations

Equations (2, 8 and 9) were solved using the finite element method by meshing the fluid domains with three-noded triangular elements. The fluid was 100 °C water in all calculations. 100 nodes were used across the lines of symmetry shown in Figure 1. This resulted in the rivulet meshes having 9,639 elements. For the films, calculations were performed for film thicknesses over the range 0 to 350 μm. For the rivulets, $\theta$ was initially set to 50º, as this was the contact angle in the falling film evaporator experiments (described in the Appendix), and calculations were performed with rivulet half angle values ($\alpha$) in the range 0 to 30º.

To calculate the velocity distribution down the tube $u_z(x,y)$, equation (2) was solved subject to the boundary conditions, $u_z = 0$ at the tube wall and $\partial u_z/\partial n_0 = 0$ at the free surface of the flow, where $n_0$ is the outer normal of the free surface.

Equation (8) for the specific turbulence kinetic energy ($k$) was solved subject to the boundary conditions $k = 0$ at the tube wall and $\partial k/\partial n_0 = 0$ at the free surface.

Equation (9) for the specific dissipation rate ($\omega$) was solved with high values of $\omega$ at both the tube wall and the liquid free surface. This ensured that the turbulent viscosity was damped out at these locations. In the viscous sublayer, near the tube wall, $v_t$ goes to zero and the production term in the $\omega$ equation can be neglected. In this region $\omega$ is approximately given by $6v/C_{\omega 2}y_p^2$ (Wilcox[31]). (This equation can be derived from the analytic solution of a one dimensional equation (9) in which the production term and $v_t$ are set to zero). In these simulations, the near wall nodes with $y^+$ values below 2.5 were set to the value $6v/C_{\omega 2}y_p^2$ and the nodes at the tube wall were set to a value of $\omega$ ten times that of the node nearest to the wall (Menter [32]).

Near the liquid free surface $v_t$ goes to zero and the production term in the $\omega$ equation is much smaller than the dissipation term and so, as in the near wall region, the value of $\omega$ increases to high values (approximately varying near the surface as the inverse square of the distance from the surface). The nodes at the liquid free surface were set to a high value of $10^6$ s$^{-1}$ in all simulations. This high value of ω was found to be effective in damping out the turbulent viscosity at the free surface. The specific dissipation rate decreases rapidly away from the tube wall and free surface, and inflation layers were incorporated into the mesh to properly model these steep gradients. Each inflation element quadrilateral was made up of two linear triangular elements.

In Case 1 for the laminar calculations, $v_t$ was set to zero at all nodes and then equation (2) was solved. In Case 2 for the $k$-$\omega$ calculations the $u_z$, $k$ and $\omega$ equations (2,8 and 9) were put in a loop and iteration was continued around the loop until convergence. In Case 3, equation (2) was solved with $v_t$ values which were calculated from the MEP equation using the iterative method described in the following steps:

1. Approximate values for the friction velocity along the tube wall were set to the values given by the equation $u^* = \sqrt{g\delta_p}$, where $\delta_p$ is the perpendicular distance from the wall to the liquid free surface

2. Using these values for the wall friction velocity, the value of $v_t$ at each of the nodes was calculated from equation (11) as described in the previous section.

3. Equation (2) was solved using these values of $v_t$ and, from the resulting velocity distribution, the wall shear stresses along the tube wall were calculated from the velocities of the nodes nearest to the wall (these nodes were always in the laminar sub layer). These values of the wall shear stress were used to calculate new values of friction velocity along the tube wall.
4. Iteration was continued around steps 2 and 3 until convergence. At convergence, the nodal values of $v_t$ calculated from equation (11) are matched with the friction velocity distribution along the tube wall.

3.4 Calculation of the film and rivulet flowrates and energies

The flowrates and energies of the flows were calculated from the nodal velocity and specific turbulence kinetic energy vectors that were output by the finite element calculations. The fluid time averaged velocity down the tube at any point in an element $u_z^{(e)}$ is given by

$$u_z^{(e)}(x,y) = N_i u_{z_i} + N_j u_{z_j} + N_k u_{z_k} \qquad (13)$$

where $N_i, N_j, N_k$ are the element shape functions for each node (labelled $i$, $j$ and $k$) and $u_{z_i}, u_{z_j}, u_{z_k}$ are the time averaged fluid velocities at each node. The film and rivulet volumetric flow rates were calculated using the equation

$$F = \sum_{e=1}^{m} \iint u_z^{(e)} dA = \frac{1}{3} \sum_{e=1}^{m} A^{(e)} [u_{z_i} + u_{z_j} + u_{z_k}] \qquad (14)$$

where $m$ is the total number of elements required to mesh the complete film or complete rivulet (a smaller number of elements being used in the calculations because of symmetry) and $A^{(e)}$ is the element area. The film flowrate $F(film)$ and the flowrate associated with one rivulet $F(riv)$ were calculated with this equation. The film and rivulet kinetic energies of the mean flow per unit length of tube were calculated using the equation

$$E_{K1} = \sum_{e=1}^{m} \iint \frac{1}{2} \rho \left[u_z^{(e)}\right]^2 dA = \frac{\rho}{12} \sum_{e=1}^{m} A^{(e)} [u_{z_i}^2 + u_{z_j}^2 + u_{z_k}^2 + u_{z_i} u_{z_j} + u_{z_j} u_{z_k} + u_{z_k} u_{z_i}] \qquad (15)$$

The derivation of equations (14) and (15) has been presented in a previous paper [8]. The kinetic energy of the mean flow per unit length of the film $E_{K1}(film)$ and the kinetic energy of the mean flow per unit length associated with one rivulet $E_{K1}(riv)$ were calculated with this equation.

Using a similar procedure to that used to evaluate $F$ in equation (14), the film and rivulet kinetic energies of the turbulence per unit length of tube were calculated from

$$E_{K2} = \sum_{e=1}^{m} \iint \rho k^{(e)} dA = \frac{\rho}{3} \sum_{e=1}^{m} A^{(e)}[k_i + k_j + k_k] \qquad (16)$$

Where $k_i$, $k_j$ and $k_k$ are the specific turbulence kinetic energies at nodes $i$, $j$ and $k$. The kinetic energy of the turbulence per unit length of the film $E_{K2}(film)$ and the kinetic energy of the turbulence per unit length associated with one rivulet $E_{K2}(riv)$ were calculated with this equation. This equation is only used with the $k$-$\omega$ model. $E_{K2}$ is zero for the laminar case and $E_{K2}$ is taken to be zero when the MEP equation is used.

Calculations were performed for films with thicknesses over the range 0 to 350 µm and for rivulets with $\alpha$ values over the range 0 to 30° (for $\theta$=50°), the calculations being performed for each of the three turbulence cases. The results of these calculations are shown in Figure 2 in which $F(film)$, $E_{K1}(film)$ and $E_{K2}(film)$ are plotted as a function film thickness and $F(riv)$, $E_{K1}(riv)$ and $E_{K2}(riv)$ are plotted as a function of the rivulet half-angle $\alpha$.

The meshes used to produce these graphs had 100 nodes along the lines of symmetry and this resulted in a mesh of 9,639 elements for the rivulet calculations. Checks were made to ensure that the film and rivulet meshes were sufficiently fine by repeating some of the calculations with 150 nodes along the lines of symmetry. This increased the number of elements in the rivulet calculations to 22,039. The difference between the values of flowrate calculated using 150 symmetry line nodes with those calculated using 100 nodes was less than 0.5 %.

3.5 Calculation of the total flowrates and energies

In this section, the energies from the previous section are combined with expressions for surface energy to give equations for the total energy per unit length of tube ($E_T$) of the flow configurations. Equations for the total flowrate ($F_T$) are also obtained.

For a film of thickness $\delta_f$ flowing down the tube, the total energy per unit length is made up of four parts. These are:

1. The kinetic energy of the mean flow $\qquad E_{K1}(film)$
2. The kinetic energy of turbulence of the flow $\qquad E_{K2}(film)$
3. The energy of the solid-liquid interface $\qquad 2\pi R \gamma_{SL}$
4. The energy of the liquid-vapour interface $\qquad 2\pi(R - \delta_f)\gamma_{LV}$

$$E_T = E_{K1}(film) + E_{K2}(film) + 2\pi R \gamma_{SL} + 2\pi(R - \delta_f)\gamma_{LV} \qquad (17)$$

which can be rewritten as

$$E_T - 2\pi R\gamma_{SL} = E_{K1}(film) + E_{K2}(film) + 2\pi(R - \delta_f)\gamma_{LV} \quad (18)$$

The left-hand side of this equation is the total energy minus a constant and is the function that is used to compare the energies of all the possible flow configurations. Also, the total flowrate is given by

$$F_T = F(film) \quad (19)$$

For $n$ identical rivulets of half-angle $\alpha$ flowing down the tube, the total energy per unit length is made up of five parts. These are:

1. The kinetic energy of the mean flow $\quad nE_{K1}(riv)$
2. The kinetic energy of the turbulence of the flow $\quad nE_{K2}(riv)$
3. The energy of the solid-liquid interface(s) $\quad 2n\alpha R\gamma_{SL}$
4. The energy of the liquid-vapour interface(s) $\quad 2n\beta S\gamma_{LV}$
5. The energy of the solid-vapour interface(s) $\quad 2[\pi - n\alpha]R\gamma_{SV}$

$$E_T = nE_{K1}(riv) + nE_{K2}(riv) + 2n\alpha R\gamma_{SL} + 2n\beta S\gamma_{LV} + 2[\pi - n\alpha]R\gamma_{SV} \quad (20)$$

Using Young's equation

$$\gamma_{SV} = \gamma_{SL} + \gamma_{LV}\cos\theta \quad (21)$$

equation (20) can be rewritten as

$$E_T - 2\pi R\gamma_{SL} = nE_{K1}(riv) + nE_{K2}(riv) + 2n\beta S\gamma_{LV} + 2[\pi - n\alpha]R\gamma_{LV}\cos\theta \quad (22)$$

Also, the total flowrate is given by

$$F_T = n\,F(riv) \quad (23)$$

## 3.6 Construction of the $E_T$ -$2\pi R\gamma_{SL}$ versus $F_T$ graph

Once expressions for the total flowrate and total energy of the film and rivulet configurations have been obtained, a total relative energy ($E_T$ -$2\pi R\gamma_{SL}$) versus total flowrate ($F_T$) graph is constructed for laminar flow. The dimensions and values of the physical properties (at 100 ºC) used in these calculations are listed in the Appendix.

In order to construct a curve for film flow on this graph, a value of $\delta_f$ in the range 0-350 μm is chosen and the corresponding values of $F$ (film), $E_{K1}$ (film) and $E_{K2}$ (film) (which is equal to zero in this laminar case) are found by interpolation of the datasets used to construct the graphs of figures 2a, 2b and 2c. The values of $F$ (film), $E_{K1}$ (film) and $E_{K2}$ (film), together with the selected value of $\delta_f$ and the known values of $R$ and $\gamma_{LV}$ are substituted into equations (18) and (19) to give values of ($E_T$ -$2\pi R\gamma_{SL}$) and $F_T$. These two values are the coordinates of a point on the energy-flowrate graph. Further values of $\delta_f$ are chosen and the energy-flowrate curve for film flow is constructed.

In order to construct a curve for rivulet flow on this graph, a value of $n$ is set and a value of $\alpha$ is chosen in the range 0-30º. Then, from the known value of $\theta$ (50º) and selected value of $\alpha$, the values of $\beta$ and $S$ are calculated from equations (4) and (5). The values of $F$ (riv), $E_{K1}$ (riv) and $E_{K2}$ (riv) (which is equal to zero in this laminar case) corresponding to the selected value of $\alpha$ are found by interpolation of the datasets used to construct the graphs of figures 2d, 2e and 2f. The values of $F$ (riv), $E_{K1}$ (riv) and $E_{K2}$ (riv), and the selected value $\alpha$, and the calculated values of $\beta$ and $S$ together with the known values of $\gamma_{LV}$, $\theta$, and $R$ and the set value of $n$ are substituted into equations (22) and (23) to give values of ($E_T$-$2\pi R\gamma_{SL}$) and $F_T$. These values are plotted as a point on the energy-flowrate graph. Further values of $\alpha$ are chosen and the energy-flowrate curve for $n$ rivulets is constructed. New values of $n$ are then set, and the corresponding energy-flowrate curves are constructed.

For laminar flow, the graph of Figure 3a results (for 100 ºC water, $R$= 3 mm and $\theta$ = 50˚). In order to make this graph easier to read, only configurations containing odd numbers of rivulets are shown. It can be seen from this graph that configurations containing more than 15 rivulets would be expected to have energies much higher than any of the lowest energy configurations over the range of flowrates considered. At a given flowrate, because of the finite circumference of the tube, together with the assumption of identical rivulet geometries, the energy levels of the system are discrete. i.e., the rivulet system does not have a continuous range of allowable energies as in the model of Mikielewicz, and Moszynski [4] in which the plane wall is effectively assumed to be infinite in extent.

For flowrates above approximately 800 mm³ s⁻¹ the film has the lowest energy and is therefore the expected configuration. As the flow is reduced from approximately 800 mm³ s⁻¹ towards zero, the number of rivulets making up the ground state configuration reduces. Graphs of the type shown in Figure 3a were also constructed for turbulent flows using the MEP and $k$-$\omega$ model data shown in the graphs of Figure 2. The results from these calculations are compared with the laminar data in Figure 3b. In the rivulet curves of this graph, only the lowest energy configurations of all the rivulet configurations have been plotted at each flowrate.

It can be seen that the incorporation of a turbulence model has given lowest energy rivulet curves which (in approximate terms) appear to be rotated clockwise from the lowest energy laminar curve. When there is turbulence in a rivulet both the total energy and total flowrate change. This can be seen from equations (22) and (23). For given point on a laminar rivulet curve with a given rivulet geometry, the presence of turbulence reduces the velocities in the rivulet, and this moves the point in the negative $x$ direction on the energy-flowrate graph. The turbulence also reduces the kinetic energy of the mean flow in the rivulet (which is the main component) and this moves the point in the negative $y$ direction on the graph. But the kinetic energies of the mean flow are proportional to the square of the velocities and so (in percentage terms) the point is moved more in the negative $y$ direction than in the negative $x$ direction. This means that the point is moved nearer the origin and clockwise about the origin. And so, when a turbulence model is used to calculate the velocity distributions, each point on each of the laminar rivulet curves either remains in the same place or is rotated clockwise and moved nearer the origin. The same effect is shown in film configurations as the films increase in thickness and a difference in energy per unit length can be seen between the laminar and $k$-$\omega$ curves at the higher flowrates of this graph. The laminar and turbulent (MEP) film curves start to differ in energy at a flowrate above that shown in this graph.

The hollow circles on each of the rivulet curves show the points at which the number of rivulets making up the lowest energy configuration change. The number of rivulets making up each configuration can be found by counting from the left-hand side of the graph. For example, for the laminar curve there is a hollow circle just below the intersection with the film curve. This circle represents the changeover between a $n$ = 5 and a $n$ = 6 rivulet configuration. And so, at the point of intersection, the lowest energy configuration is made up of six rivulets. For the MEP curve, there also is a hollow circle just below the intersection with the film curve. This circle represents the changeover between a $n$ = 2 and a $n$ = 3 configuration, And, therefore, at the point of intersection the lowest energy configuration is made up of 3 rivulets.

From graphs 3a and 3b, it can be seen how this numerical calculation procedure is related to the model of Mickielewicz and Moszynski [4]. In the present calculation, the rivulet energy curves are plotted on a graph and at a given flowrate the lowest energy configuration is found by inspection. This gives a curve of lowest rivulet energy at each flowrate. The intersection of this curve with the film energy curve gives the point below which a film is not expected to be stable. In the analytic model of Mikielewicz and Moszynski [4], expressions for the film and rivulet mass flows and energies are equated, and the stable rivulet configuration is found by differentiation.

From Fig 3b it can be seen that incorporation of the effect of turbulence into the calculations has increased the values of the critical flowrate $F_*$ (the flowrate below which a film is not expected to be stable). The values of $F_*$ are: $F_*$(Laminar)= 765 mm$^3$ s$^{-1}$, $F_*$(MEP)= 853 mm$^3$ s$^{-1}$ and $F_*(k\text{-}\omega)$ = 1088 mm$^3$ s$^{-1}$. The $k$-$\omega$ model gives a higher value of $F_*$ than the MEP equation because the level of turbulence is higher in the $k$-$\omega$ rivulets. This is shown, for example, by the turbulent viscosity contours of an $\alpha$ = 18.5º rivulet (Fig 4). This rivulet geometry has been chosen because at a flowrate just below the critical flowrate of 1088 mm$^3$ s$^{-1}$ for the $k$-$\omega$ model, the film is predicted to break up into 3 x 18.5º rivulets. In this figure the viscosity contours calculated by both the $k$-$\omega$ model and by the MEP equation are shown. The $k$-$\omega$ contours extend further away from the centreline than the MEP contours.

As a first approximation, the rivulet can be thought of as being made up of a number of radial strips, with the thickness of the strips increasing towards the centreline. If these strips were to have the same velocity profiles as those modelled in the film calculations, then these strips would have properties similar to those shown in Fig 2a. In this graph for film flow, the film thickness at which the laminar and turbulent $k$-$\omega$ curves diverge is lower than the thickness at which the laminar and turbulent (MEP) curves diverge. The laminar and turbulent($k$-$\omega$) curves start to diverge at a flowrate of about 1340 mm$^3$ s$^{-1}$ which corresponds to a Reynolds number of approximately 960, which is far too low. However, as the film thickness increases, the agreement between the $k$-$\omega$ model and the MEP equation is much closer. This can be seen indirectly from the closeness of the $k$-$\omega$ and MEP curves in Fig 2d at higher values of $\alpha$. Kharangate et al. [30] who studied falling films with higher film thicknesses also found good agreement between the MEP equation and the $k$-$\omega$ model.

Turbulence is predicted to start at lower Reynolds numbers and film thicknesses when the $k$-$\omega$ model is used, and this is consistent with the broadening of the $k$-$\omega$ model contours in the rivulet. It is concluded that, since the $k$-$\omega$ model predicts the onset of turbulence at Reynolds number that are too low in the falling film simulations, the model overestimates the level of turbulence within the rivulets.

The calculations presented here have shown that whatever turbulence model is used, it must be able to give good predictions in the transition regime. The MEP equation has been shown to give good agreement with experiments and can be used in the transition regime, and the MEP model is the model that is used from this point on. The incorporation of an intermittency factor into the $k$-$\omega$ model to improve predictions in the transition regime is discussed in section 6.

3.7 The effect of contact angle on the critical film thickness

A graph of the type shown in Fig. 3b was constructed for three further water-tube wall contact angles ($\theta$ = 30º, 70º and 90º) for both the laminar and turbulent (MEP) cases. From these graphs the variation of the critical film thickness $\delta_*$ with $\theta$ was found. The results of these calculations are shown in Table 2 together with the rivulet geometries that would be expected to exist if the flow were reduced to just below the critical flowrate. In this table the critical values are shown for film thickness ($\delta_*$), flowrate ($F_*$), number of rivulets ($n_*$), rivulet half angle ($\alpha_*$), rivulet centreline thickness ($\delta_{max_*}$)(Fig 1) and wetted to total tube area ($X_*$).

These data show that the critical flowrates increase as the contact angles increase. For films, the increase in flowrate is associated with an increase in the critical film thicknesses. For the rivulets, the wetted to total tube area does not change much and the increase in flowrate is provided by an increase in the centreline thicknesses ($\delta_{max_*}$) of the rivulets.

In section 3.2 the film thickness at which turbulence is first expected to appear was calculated to be 226 µm (for $R$ = 3 mm). The values of critical rivulet centreline thickness ($\delta_{max_*}$) shown in Table 2 for contact angles above 30º in the laminar calculation are all above this value. And so, for these cases, the laminar calculation has predicted laminar rivulet geometries that, based on the model presented in this paper, would not be able to exist in a laminar condition. These systems are modified as shown in the adjacent turbulent(MEP) column of the table.

In Figure 5, these critical film thicknesses are compared with the values predicted by the laminar plane wall minimum total energy (MTE) model of Mikielewicz and Moszynski [4] and the laminar plane wall force balance (FB) model of Hartley and Murgatroyd [1]. A calculation to estimate the effect of turbulence on the FB model is presented in the Appendix. It is concluded from this calculation that the effect of turbulence on the FB model is small. The curve for the FB model in which turbulence is incorporated is very close to the laminar FB curve shown in Figure 5 and so it has not been included on the graph.

The data shown in Figure 5 show that inclusion of the effects of turbulence in the MTE model can significantly increase the values of the critical film thickness. The effect of turbulence increases as the tube to liquid contact angle increases. The results from the laminar model are close to the values predicted by [4]. There are three reasons why the laminar flow results presented here differ from those of the plane wall model of Mikielewicz and Moszynski [4]. These are:

1.The flow in reference [4] is down a plane wall whereas the flow in the laminar model in this paper is down a tube.

2.In reference [4], the velocity distribution in a rivulet is calculated by dividing the rivulet into thin strips and then assuming that the velocity profile in each strip is the same as the velocity distribution in a uniform film of the same thickness. In the laminar model of this paper the velocity distributions are numerically exact.

3.In reference [4], the plane wall is effectively assumed to be infinite in extent and the energy levels form a continuum whereas in the model presented in this paper the tube circumference is finite and, with the assumption of identical rivulet geometries, the system does not have a continuous range of allowable energies at a given flowrate. A generalised method of calculation in which non identical rivulet geometries are considered is presented in [8].

## 4  Heat Transfer Calculations

Once the fluid geometries that are expected to exist at each flowrate have been calculated, the associated heat transfer coefficients for the configurations are calculated using the equation

$$\frac{\partial}{\partial x}\left(\lambda' \frac{\partial T}{\partial x}\right) + \frac{\partial}{\partial y}\left(\lambda' \frac{\partial T}{\partial y}\right) = 0 \qquad (24)$$

where $\lambda'$ is the thermal conductivity and is given by

$\lambda' = \lambda_S$ in the (solid) tube wall, and

$\lambda' = \rho C_p \left(\frac{\nu}{Pr} + \frac{\nu_t}{Pr_t}\right)$ in the fluid

The fluid and solid domains are meshed with triangular elements and equation (24) is solved using the finite element method.  The meshes are the same as those used in the fluid flow calculations but with an additional region to model the tube wall.  The turbulent viscosities ($\nu_t$) which were calculated in the fluid flow calculations were passed to the thermal calculations. The dimensions and physical properties of this system are listed in the Appendix. In the film calculations, the mesh comprises a tube wall of thermal conductivity $\lambda_S$ which spans the radial distances $R$ and $R + \delta_S$ and a falling film which extends from $R - \delta_f$ to $R$.  In the rivulet flow calculations, the mesh comprises a tube wall of thermal conductivity $\lambda_S$ which spans the radial distances $R$ and $R + \delta_S$ and half a rivulet which subtends an angle $\alpha$ at the tube axis.

The nodes on the fluid free surface are fixed at a temperature $T_0$ (100 ºC) and a condensing heat transfer coefficient $h_{ext}$ is applied on the outer surface of the tube wall at $R + \delta_S$ with an associated external temperature of $T_0 + \Delta T$ ($\Delta T$ = 0.52 ºC).    For the plastic tube evaporator presently under consideration, $R$ = 3 mm, $\lambda_S$ = 0.3 W m⁻¹ K⁻¹ and the water thermal conductivity $\lambda$ = 0.681 W m⁻¹ K⁻¹.  $Pr$ was 1.75 and $Pr_t$ was assumed to be equal to 0.8. The condensation outside the tube was dropwise and an appropriate value for the external heat transfer coefficient $h_{ext}$ is approximately 55 kW m⁻² K⁻¹. Calculations were performed for films with thicknesses over the range 0 to 350 $\mu$m and for rivulets with values of the half angle $\alpha$ over the range 0 to 30º (with $\theta$ set to 50º), the calculations being performed for both the laminar case and the turbulent case described by the MEP equation. From the nodal temperature solution vectors, the rates of heat flow through the wetted parts of the tube wall were calculated. The results of these calculations are shown in figures 6a and 6b. The heat transfer rate per unit length of tube for a complete film $Q$ *(film)* is plotted as a function of the film thickness ($\delta_f$) and the heat transfer rate per unit length of tube for a single rivulet $Q$ *(riv)* is plotted as a function of the rivulet half-angle ($\alpha$).

The heat transfer performance of the whole tube was obtained by defining an overall heat transfer coefficient for the system ($U$) in terms of the inside area of the tube, such that:

For film configurations $\qquad U = \dfrac{Q(film)}{2\pi R \Delta T} \qquad (25)$

For rivulet configurations $\qquad U = \dfrac{nQ(riv)}{2\pi R \Delta T} \qquad (26)$

The change in heat transfer performance of the plastic tube evaporator was obtained by first finding the geometries of the lowest energy configurations (in terms of $\delta_f$ and $\alpha$) that were expected to exist at each flowrate. The values of $\delta_f$ or $\alpha$ (and the number of rivulets) associated with each lowest energy configuration were found from the data used to construct the energy flowrate graph of Fig 3b.  Once the geometries of the flows were established, the associated heat transfer properties were calculated using equations (25) and (26) together with the data from Figures 6a and 6b. The results from these calculations are shown in Fig. 6c together with the experimental results from the evaporator.
.
5  Comparison of model predictions with experimental results

From the graph of Fig 6c it can be seen that there is good agreement between experiment and theory for film flows above about  2000 mm³s⁻¹.  The model also correctly predicts a reduction in heat transfer coefficient as the flow reduces to zero.  The experimental coefficients display a gradual decrease to zero whereas the calculated coefficients for both the laminar and turbulent cases display a step change at critical flowrates below 1000 mm³s⁻¹.  However, the inclusion of turbulence in the MTE model moves the critical flowrate nearer to the point at which the experimental results reduce from the film value.

In the film flow region of the graph, the laminar model incorrectly predicts a continued decrease in coefficient as the flowrate increases. In the turbulent calculation the coefficient curve correctly shows a minimum at a Reynolds number of approximately 1800.
The turbulent calculations (shown in Table 2) predict rivulet geometries that are significantly different from those predicted in laminar calculations at the point of film breakup. In the laminar calculation, at a flowrate just below 765 mm$^3$ s$^{-1}$, the film breaks up into 6 x 11.9⁰ rivulets ($X_*$ = 0.4). In the turbulent calculation, at a flowrate just below 853 mm$^3$ s$^{-1}$, the film breaks up into 3 x 15.7⁰ rivulets ($X_*$ = 0.26). Although $X_*$ (the wetted to total tube area at breakup) for the turbulent calculation is less than that of the laminar case, the heat transfer through the rivulets is increased by the increase in heat transfer provided by the turbulence

6. Discussion

The effect of turbulence has been incorporated into a model in which the minimum film thicknesses for flows down a vertical tube were calculated by comparing the total energies of films with the total energies of a number of rivulets. The velocity distributions of the flows were calculated using the finite element method. The finite element programs for the fluid flow and thermal calculations were written in C.

The main conclusion from the calculations presented in this paper is that inclusion of the effect of turbulence in the minimum total energy (MTE) model can significantly increase the predicted values of the minimum film thickness. The effect of turbulence was found to increase as the contact angle increased. For a flow at a contact angle of 30⁰ there was no increase in $\delta_*$, but at a contact angle of 90⁰, $\delta_*$ increased from 201 μm to 245 μm and the corresponding increase in the critical flowrate ($F_*$) was from 1584 mm$^3$ s$^{-1}$ to 2773 mm$^3$ s$^{-1}$. At the highest contact angle of 90⁰, therefore, the predicted critical film thickness increased by 22 % and the predicted critical flowrate increased by 75 %. Also, incorporation of turbulence into the MTE model significantly reduced the number of rivulets in the rivulet configuration at the minimum flowrate. At a contact angle of 50⁰, the number of rivulets at the critical flowrate was predicted to be 6 in the laminar case but this reduced to 3 when a turbulence calculation was used

The inclusion of turbulence into the force balance model (FB) also increased the values of the minimum film thickness, although effect was much less than that found in the MTE calculations. In the FB model the effect of turbulence started at a contact angle of approximately 70⁰ and at an angle of 90⁰ the critical film thickness increased from the laminar value by approximately 1 %. The reason for this difference is that the FB model only considers the film, but the MTE model considers both the film and rivulet configurations and, at the critical conditions described by the MTE model (as shown in Table 2), the centreline thicknesses of the rivulets ($\delta_{max_*}$) are approximately twice those of the associated values of $\delta_*$. And so, these rivulet configurations are more likely to be affected by the introduction of a turbulence model into the calculations.

The results from the MTE calculations were used to model the thermal performance of a falling film evaporator and the model results were compared to data obtained from the evaporator experiments. There was good agreement between experiment and the MTE model for film flows above a Reynolds number of approximately 1500. The model also correctly predicted a reduction in heat transfer coefficient as the flow reduces to zero. It would have been a considerable advantage if the film and rivulet flows in the evaporator experiments could have been visually observed. This would have allowed comparison of the rivulet geometries with those predicted by the MTE model. However, even though the plastic tube was thin walled and opaque, it was not possible to gain a clear impression of the type of flow inside the tube

The inclusion of turbulence into the model is a logical way to develop the model because, as it was shown in section 3.7, the values of critical rivulet centreline thickness ($\delta_{max_*}$) for the laminar MTE model (Table 2) are above the value of $\delta_{turb}$ (the film thickness at which the film would be expected to become turbulent) when the contact angle is above 30º. And so, for these cases, the laminar calculation has predicted laminar rivulet geometries that, based on the model presented in this paper, would not be expected to be able to exist in a laminar condition.

The calculations for the MTE model have shown that whatever turbulence model is used, it must be able to give good predictions in the transition regime. The turbulence calculations were performed using two methods: a $k$-$\omega$ turbulence model and the Mudawar and El-Masri turbulent viscosity profile (MEP). Both methods have advantages and disadvantages. The advantage of the $k$-$\omega$ model is that it is suitable for use in a 2D geometry, the disadvantage is that it predicts the onset of turbulence at Reynolds numbers that are too low. The advantage of the MEP equation is that it shows good agreement with experimental data and can be used in the transition region , but the disadvantage is that it is a 1D profile and an approximation has to be made in order to apply it to the 2D rivulet case. However, the approximation that is made in applying the 1D MEP profile to the 2D rivulet case is similar to the approximation that is made in the Mikielewicz and Moszynski model [4] to obtain the velocity profile in a rivulet. It should be noted that application of a 1D profile to a 2D rivulet geometry may overestimate the turbulence in the rivulet as there is diffusion of turbulence kinetic energy away from the rivulet centre to the sides of the rivulet where the flow is always laminar. Further work is required to investigate the accuracy of applying the 1D profile to the 2D rivulet geometry as it has been in these calculations.

Ideally, for MTE calculations, a modified $k$-$\omega$ model would be used which has good agreement with experimental data in the transition region. Abraham, Sparrow, Gorman, Zhao and Minkowycz [33] have developed an intermittency model for laminar, transitional, and turbulent internal flows. This model has been successfully applied to flows with Reynolds numbers that ranged from 100 to 100,000 in circular tubes and parallel plate channels. The results from this intermittency model were found to be highly dependent on the upstream flow conditions in the transition regime. If, therefore, an intermittency model of this type were used as part of a minimum thickness calculation, the upstream flow conditions would need to be specified in the simulations, and the simulations would have to be three dimensional.

In the models presented in this paper the turbulence has been assumed to be only a function of the flow geometry whereas it is known that transitional turbulence, depends on a number of factors amongst which are surface roughness, upstream disturbance and upstream turbulence. In addition, waves are known to be present on the surface of falling films and rivulets and this has not been included in the models.

However, the structure of the MTE model presented here is such that the effect of surface waves can be incorporated into the model. This would involve performing 3D simulations for falling films and rivulets of different sizes in which the hydrodynamic character of the wavy liquid-vapour interface is computed (as in ref [16]). The results from these simulations would then be incorporated into graphs of the type shown in Fig 2 of the present paper. The model could also be extended to incorporate the effect of vapour shear.

There is the question of whether the overall approach used in these calculations is reasonable, that is, whether it is appropriate to calculate $\delta_*$ based on the comparison of the energies of fully developed film flows with the energies of the fully developed flows of a number of rivulets. Further work is needed to find out whether calculation of critical film thicknesses based on this type of comparison is appropriate.

Appendix A - Description of the plastic tube evaporator

The plastic tube evaporator consisted of a 4.15 m long vertical nylon-12 tube of internal diameter 6 mm and wall thickness 150 µm surrounded by a glass shell. Steam held at approximately 2 kPa above atmospheric pressure condensed outside the plastic tube and the heat released from this condensation travelled through the plastic wall to evaporate a small proportion of the demineralised water flowing down the inside of the tube. The condensation outside the tube was dropwise. The absolute pressure at the tube outlet was atmospheric.

The tube side water was distributed evenly around the top of the inside of the plastic tube. At the base of the tube, the steam produced from the evaporation was separated from the tube side water and this steam was passed through a condenser. In each experiment, the flowrates of both the tube side water and the condensed steam were recorded. The evaporator duty was calculated from the rate of collection of the condensed steam. In each experiment, an overall heat-transfer coefficient was calculated from the duty, overall temperature difference (0.52 ºC) and the total tube area available for heat transfer.

In the experiments, the amount of water evaporated in the tube was always less than 2.5 % of the flow introduced at the top of the tube, and at the highest experimental flowrate, the film Reynolds number was approximately 2300. The contact angle between the nylon tube and the water during the heat-transfer experiments was approximately 50º. Further details of this experiment are presented in [26].

Appendix B - Dimensions and Physical Properties

The constants, dimensions and values of the physical properties (at 100ºC) used in these calculations are:

$C_p$ = 4217 J kg$^{-1}$ K$^{-1}$
$g$ = 9.81 m s$^{-2}$
$h_{ext}$ = 55 kW m$^{-2}$ K$^{-1}$
$R$ = 3 mm
$T_0$ = 100 ºC
$\gamma_{LV}$ = 58.78 mN m$^{-1}$
$\delta_S$ = 150 µm
$\Delta T$ = 0.52 ºC
$\lambda$ = 0.681 W m$^{-1}$ K$^{-1}$
$\lambda_S$ = 0.3 W m$^{-1}$ K$^{-1}$
$\mu$ = 283.1 µN s m$^{-2}$
$\rho$ = 958.12 kg m$^{-3}$

From which
$\Lambda = (\rho^3 g^2 / 15 \mu^2 \gamma_{LV})^{1/5}$ = 4127 m$^{-1}$
$\nu = \mu/\rho$ = 2.95 x10$^{-7}$ m$^2$ s$^{-1}$
$Pr = C_p \mu / \lambda = 1.75$
$Pr_t$ is assumed to be equal to 0.8

Also, for equation (11)
$K$ = 0.4
$Ka$ = $\mu^4 g / \rho \gamma_{LV}^3$
$Re_{crit}$ = $0.04/Ka^{0.37}$ = 1672 (for evaporation [11])

# Appendix C – Approximate calculation using the force balance model

In the model of Hartley and Murgatroyd [1], the minimum thickness of a liquid film flowing down a vertical plate under the action of gravity is found from a force balance at the upstream stagnation point of a dry patch such that

$$dw \int_0^{\delta_f} \frac{\rho}{2}[u_z(y_p)]^2 \, dy_p = \gamma_{LV}(1 - \cos\theta) \, dw \qquad (27)$$

where $u_z(y_p)$ is the velocity of the fluid down the plate as a function of the perpendicular distance from the plate $y_p$, and $dw$ is a small distance across the width of the liquid stream at the stagnation point. The right-hand side of this equation is the restraining force due to surface tension. This equation can be used for laminar and turbulent flows. When a laminar velocity profile is put into equation (27) the minimum film thickness is found to be related to the contact angle by equation (1) where $\Lambda$ has the value 4127 m$^{-1}$ for 100 ºC water.

In section 3.2 the film thickness at which a 100 ºC water film flowing down a plane wall would be expected to become turbulent ($\delta_{turb}$) was calculated as 223 µm. When this value of film thickness is substituted into equation (1), $\theta$ is calculated to be 70º. And so, when $\theta$ is greater than 70º, the minimum film thicknesses will have started to become turbulent. At an angle of 90º, the minimum film thickness calculated from equation (1) is 242 µm. And therefore, when the contact angle is 90º the surface tension forces at the stagnation point are balanced by the force that would have been provided a laminar film of 242 microns.

The curves shown in the graph of $E_{k1}(film)$ of Fig 2b have been calculated by evaluating the integral $\iint \frac{1}{2}\rho[u_z]^2 dA$ for film flows down a tube of radius $R$. If the term on the left-hand side of equation (27) is denoted by $\Phi(\delta_f)$, then the values calculated by the function $E_{k1}(film)$ will be proportional to the values calculated by the function $\Phi(\delta_f)$ when $\delta_f$ is small compared to the tube radius $R$ ($R$ = 3 mm being used in the $E_{k1}(film)$ calculations). It is found from the laminar and turbulent (MEP) data used to construct graph of Figure 2b that a laminar film of 242 microns has the same value of $E_{k1}(film)$ as a 243.5 micron film calculated using the turbulent(MEP) equation. If the proportionality of the two functions ($E_{k1}(film)$ and $\Phi(\delta_f)$) is assumed to extend to 250 microns, then the value of $\Phi(\delta_f)$ calculated for a 242 micron laminar film would equal the value of $\Phi(\delta_f)$ calculated for a 243.5 micron film in which the MEP equation was used. And so, when the contact angle is 90º, the force required to balance the surface tension forces (which has been calculated to be equivalent to that which could be supplied by a laminar 242 micron film) can be supplied by a turbulent(MEP) film of thickness of 243.5 microns. Although this calculation can only be considered to be approximate, it is concluded that the effect of turbulence on the minimum thicknesses calculated using a force balance model on this system is small. The values of $\delta_*$ obtained from a turbulent calculation would start to increase above the laminar values at a contact angle of 70º, and at an angle of 90º the value of $\delta_*$ would increase from 242 to approximately 243.5 microns, an increase of approximately 1 %.

REFERENCES

[1] D. E. Hartley and W.Murgatroyd, Criteria for the break-up of thin liquid layers flowing isothermally over solid surfaces, International Journal of Heat and Mass Transfer 7 (1964) 1003-1015. https://doi.org/10.1016/0017-9310(64)90042-0
[2] T.Hobler, Minimal surface wetting, Chemia Stosow 2B (1964) 145-149 (in Polish).
[3] S.G. Bankhoff, Minimum thickness of a draining liquid film, International Journal of Heat and Mass Transfer 14 (1971) 2143-2146. https://doi.org/10.1016/0017-9310(71)90034-2
[4] J. Mikielewicz and J. R.Moszynski, Minimum thickness of a liquid film flowing vertically down a solid surface, International Journal of Heat and Mass Transfer 19 (1976) 771-776
https://doi.org/10.1016/0017-9310(76)90130-7
[5] H.H. Saber and M.S El-Genk, On the breakup of a thin liquid film subject to interfacial shear, Journal of Fluid Mechanics 500 (2004) 113–133. DOI:10.1017/S0022112003007080
[6] S.K.Wilson, J.M Sullivan and B.R. Duffy, The energetics of the breakup of a sheet and of a rivulet on a vertical substrate in the presence of a uniform surface shear stress, Journal of Fluid Mechanics 674 (2011) 281-306. DOI: https://doi.org/10.1017/S0022112010006518
[7] Junqing Wei, Jinping Liu, and Xiongwen Xu, Theoretical and Experimental Investigation of the Minimum Wetting Rate of a Falling Water Film on the Vertical Grooved Plate, Ind. Eng. Chem. Res. 61 (2022) 845−854. https://doi.org/10.1021/acs.iecr.1c02533
[8] D.T.Hughes and T.R. Bott, Minimum thickness of a liquid film flowing down a vertical tube, International Journal of Heat and Mass Transfer 41 (1998) 253-260.
DOI:10.1016/S0017-9310(97)00151-8
[9] P.Schmuki and M. Laso, On the stability of rivulet flow, Journal of Fluid Mechanics 215 (1990) 125 – 143. DOI: https://doi.org/10.1017/S0022112090002580
[10] T.A. Ghezzehei, Constraints for flow regimes on smooth fracture surfaces, Water Resources Research 40 (2004) W11503, doi:10.1029/2004WR003164
[11] I.A Mudawar and M.A.El-Masri, Momentum, and heat transfer across freely falling turbulent liquid films, International Journal of Multiphase Flow 12(5) (1986) 771-790
https://doi.org/10.1016/0301-9322(86)90051-0
[12] R. F. Allen and C. M.Biggin, Longitudinal flow of a lenticular liquid filament down an inclined plane, Physics of Fluids 17 (1974) 287-291. https://doi.org/10.1063/1.1694713

[13] C.A.Perazzo and J. Gratton, Navier-Stokes solutions for parallel flow in rivulets on an inclined plane, Journal of Fluid Mechanics 507 (2004) 367–379. https://doi.org/10.1017/S0022112004008791

[14] F.H.H.Al Mukahal, B. R. Duffy, S. K. Wilson, A rivulet of a power-law fluid with constant contact angle draining down a slowly varying substrate, Physics of Fluids 27 (2015) 052101. https://doi.org/10.1063/1.4919342

[15] A. v. B. Lopes, R. M. Borges, G. C. Matias, B. G. Pimenta, I. R. Siqueira, On lubrication models for vertical rivulet flows, Meccanica 57 (2022) 1071–1082. https://doi.org/10.1007/s11012-022-01503-x

[16] N. Mascarenhas and I.Mudawar, Study of the influence of interfacial waves on heat transfer in turbulent falling films, International Journal of Heat and Mass Transfer 67 (2013) 1106-1121. https://doi.org/10.1016/j.ijheatmasstransfer.2013.08.100

[17] S. V. Alekseenko, A. V. Bobylev, V. V. Guzanov, D. M. Markovich, S. M. Kharlamov, Regular waves on vertical falling rivulets at different wetting contact angles, Thermophysics and Aeromechanics 17(3) (2010) 345-357. DOI:10.1134/S0869864310030054

[18] K.R. Morison, Q.A.G. Worth, N.P. O'dea, Minimum Wetting and Distribution Rates in Falling Film Evaporators, Food and Bioproducts Processing 84 (2006) 302-310. https://doi.org/10.1205/fbp06031

[19] Y. Lu, F. Stehmann, S. Yuan, S. Scholl, Falling film on a vertical flat plate – Influence of liquid distribution and fluid properties on wetting behavior, Applied Thermal Engineering 123 (2017) 1386-1395. https://doi.org/10.1016/j.applthermaleng.2017.05.110

[20] C. Dreiser and H. Bart, Falling film break-up and thermal performance of thin polymer film heat exchangers, International Journal of Thermal Sciences 100 (2016) 138-144. https://doi.org/10.1016/j.ijthermalsci.2015.09.022

[21] S. Paramalingam, J.Winchester, C.Marsh, On the Fouling of Falling Film Evaporators Due to Film Break-Up, Food and Bioproducts Processing 78 (2000) 79-84. https://doi.org/10.1205/096030800532770

[22] T.Munakata, K.Watanabe, K.Miyashita, Minimum wetting rate on wetted-wall column - Correlation over wide Range of liquid viscosity, Journal of Chemical Engineering of Japan, 8 (1975) 440-444. https://doi.org/10.1252/jcej.8.440

[23] K.R. Morrison and G.Tandon, Minimum Wetting Rates for Falling films on Stainless Steel, Developments in Chemical Engineering and Mineral Processing, 14 (2006) 153-162. https://doi.org/10.1002/apj.5500140113

[24] A.I.Fedorchenko and J.Hruby, On formation of dry spots in heated liquid films. Physics of Fluids 33 (2021) 023601. DOI:10.1063/5.0035547

[25] A.I.Fedorchenko and R.A.Abdulkhalikov, Metastable flow regimes of a thin liquid film on a vertical surface, Thermophysics and Aeromechanics 6(3) 1999 379-382.

[26] D.T.Hughes and T.R.Bott, The breakup of falling films inside small diameter tubes, Chemical Engineering Science 46(7) (1991) 1795-1805. https://doi.org/10.1016/0009-2509(91)87026-9

[27] D.T.Hughes, T.R.Bott and D.C.F.Pratt, Plastic tube heat transfer surfaces in falling film evaporators, 2nd U.K. National Heat Transfer Conference, Glasgow (1988) 1101-1113.

[28] D.C.F.Pratt, The Courtaulds desalination process, Desalination 42 (1982) 37-45. https://doi.org/10.1016/S0011-9164(00)88739-1

[29] D. C. F.Pratt, Evaporators, European Patent EP0159885 A2, 1985.

[30] C.R.Kharangate, H.Lee, I.A.Mudawar, Computational modeling of turbulent evaporating falling films, International Journal of Heat and Mass Transfer, 81 (2015) 52-62. http://dx.doi.org/10.1016/j.ijheatmasstransfer.2014.09.068

[31] D.C.Wilcox, Turbulence Modeling for CFD, 3rd edition, DCW Industries, Inc., La Canada CA, 2006

[32] F.R. Menter, Zonal Two Equation k-ω Turbulence Models for Aerodynamic Flows. AIAA Journal (1993) AIAA-93-2906. https://arc.aiaa.org/doi/10.2514/6.1993-2906

[33] J. P. Abraham, E. M. Sparrow, J. M. Gorman, Yu Zhao, W. J. Minkowycz, Application of an Intermittency Model for Laminar, Transitional, and Turbulent Internal Flows, Journal of Fluids Engineering 141(7) (2019) 071204.
DOI: 10.1115/1.404266

Figures and Tables

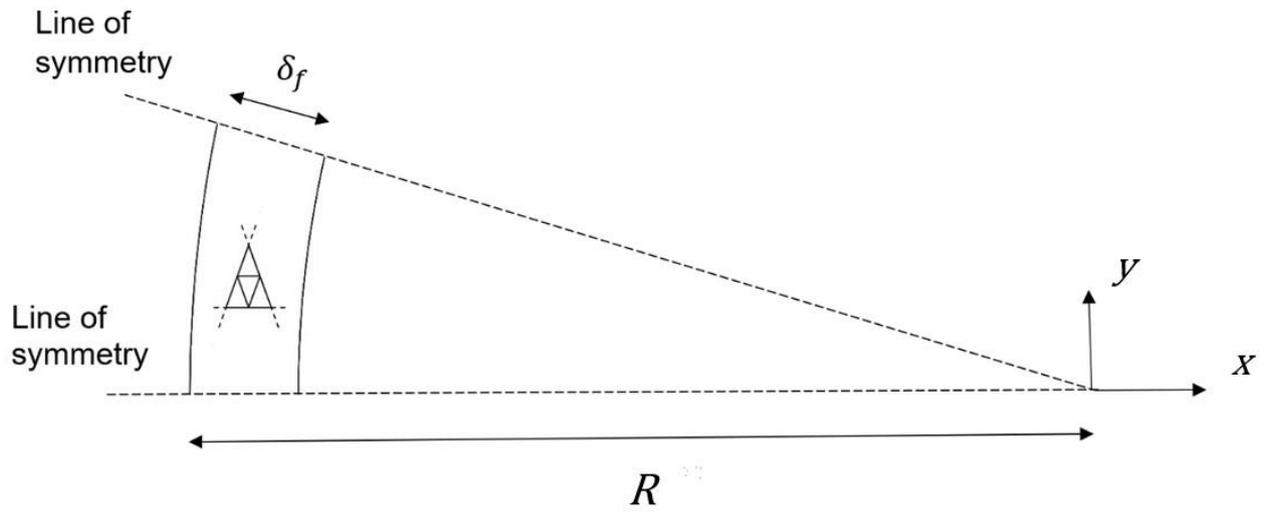

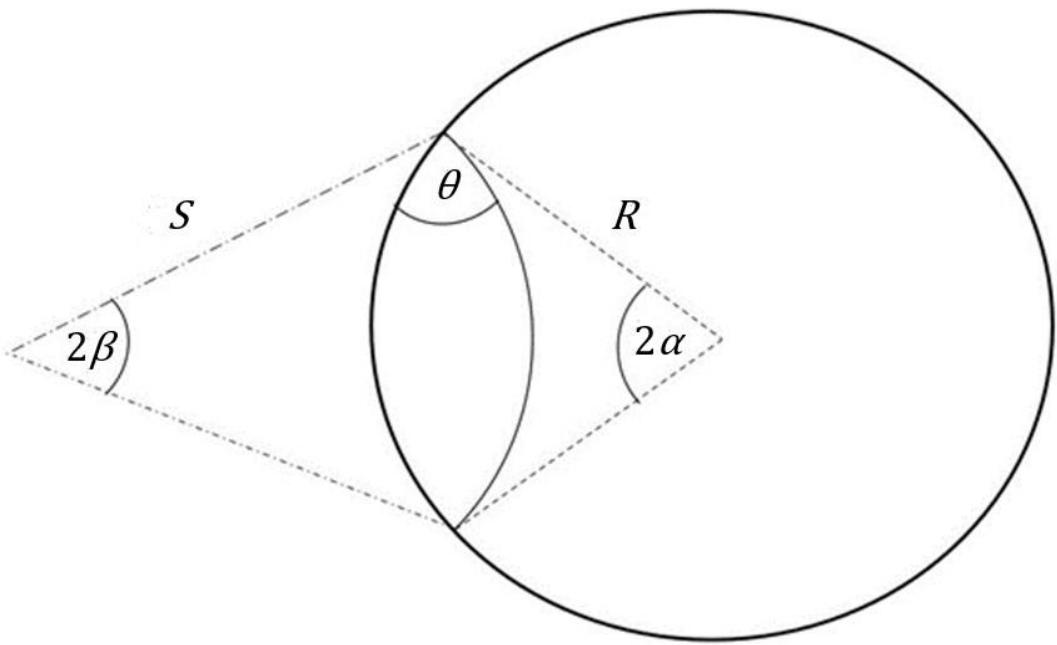

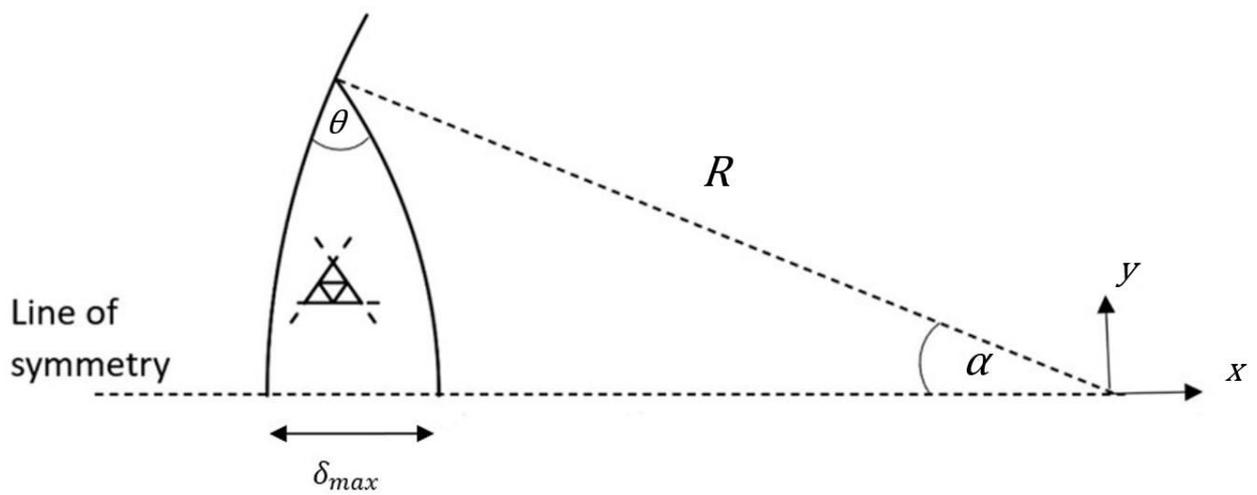

Fig. 1. Film and Rivulet Geometries

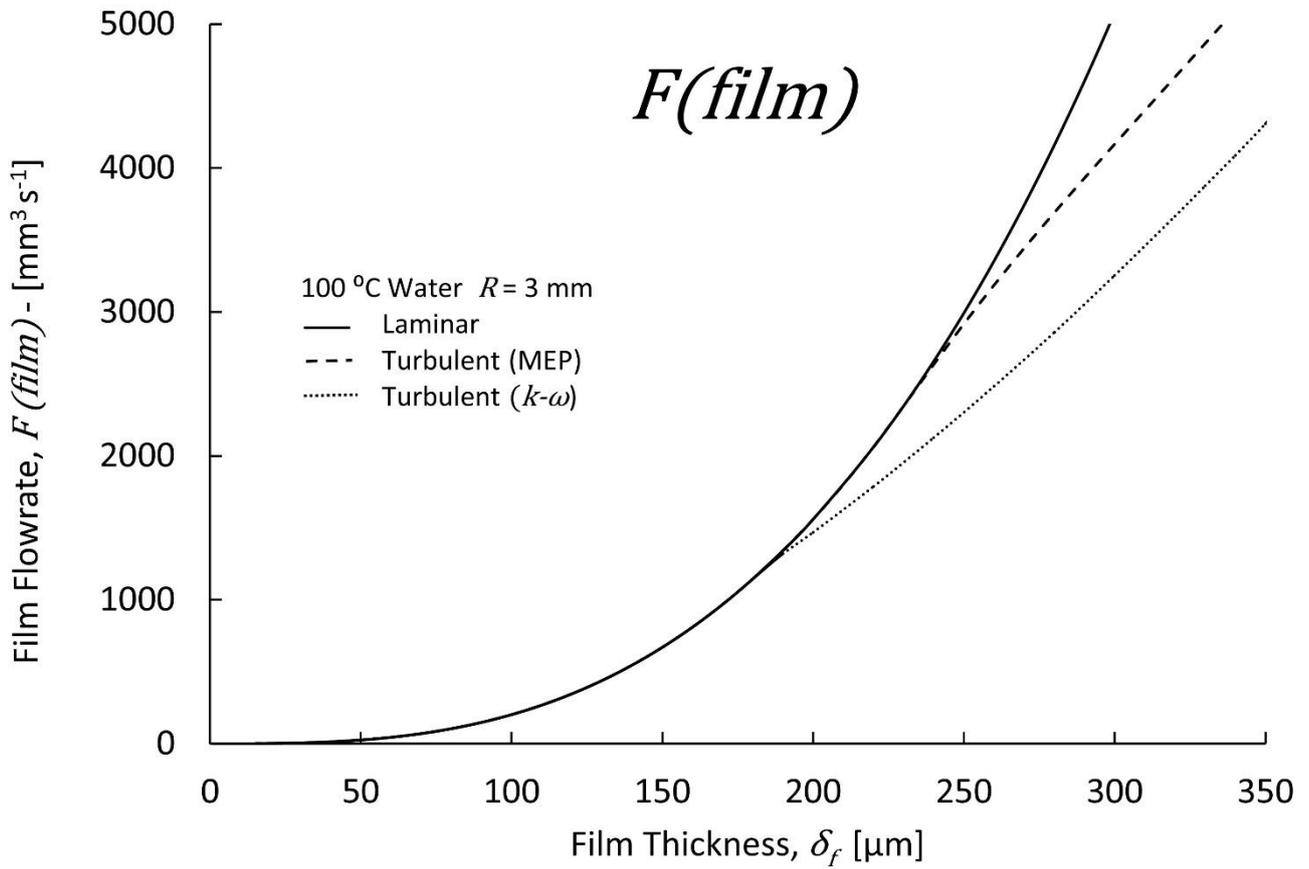

Fig. 2a  Film flowrate as a function of film thickness

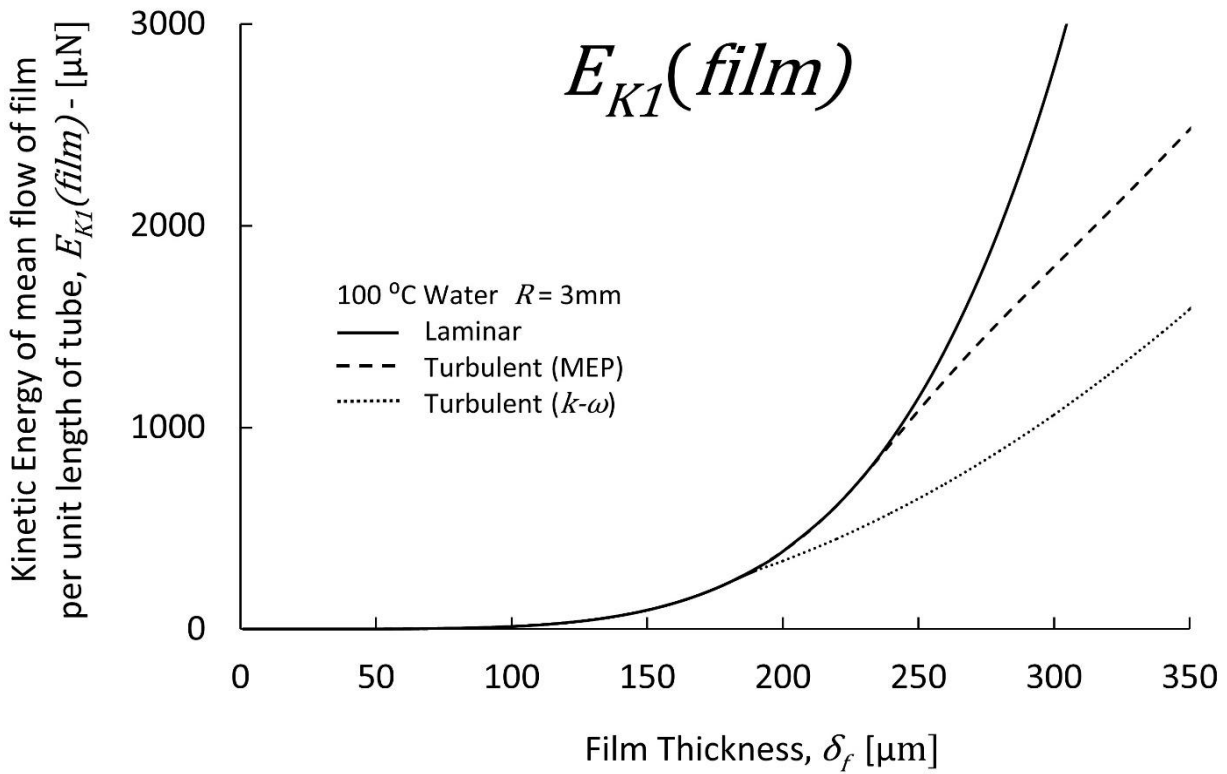

Fig. 2b  Kinetic energy of the mean flow as a function of film thickness

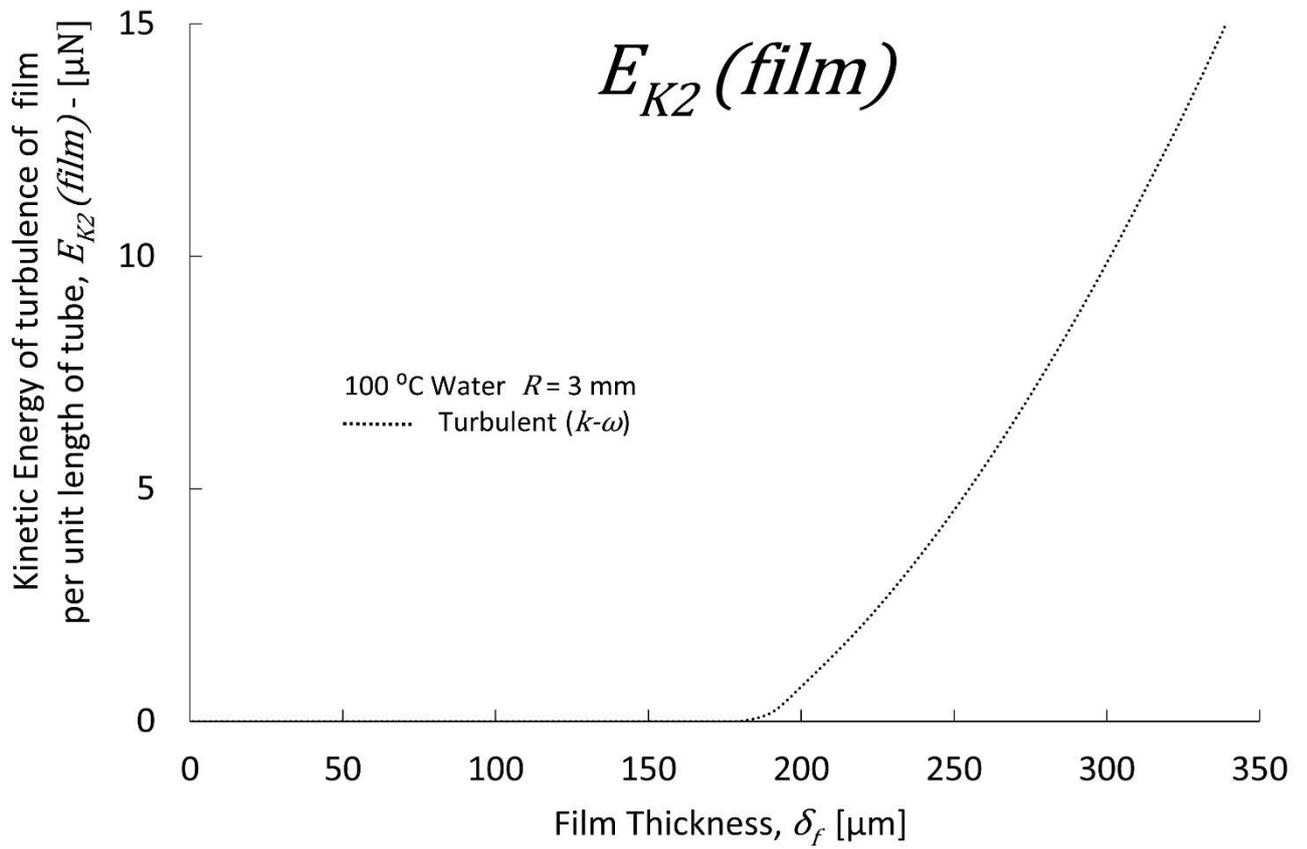

Fig. 2c   Kinetic energy of turbulence as a function of film thickness

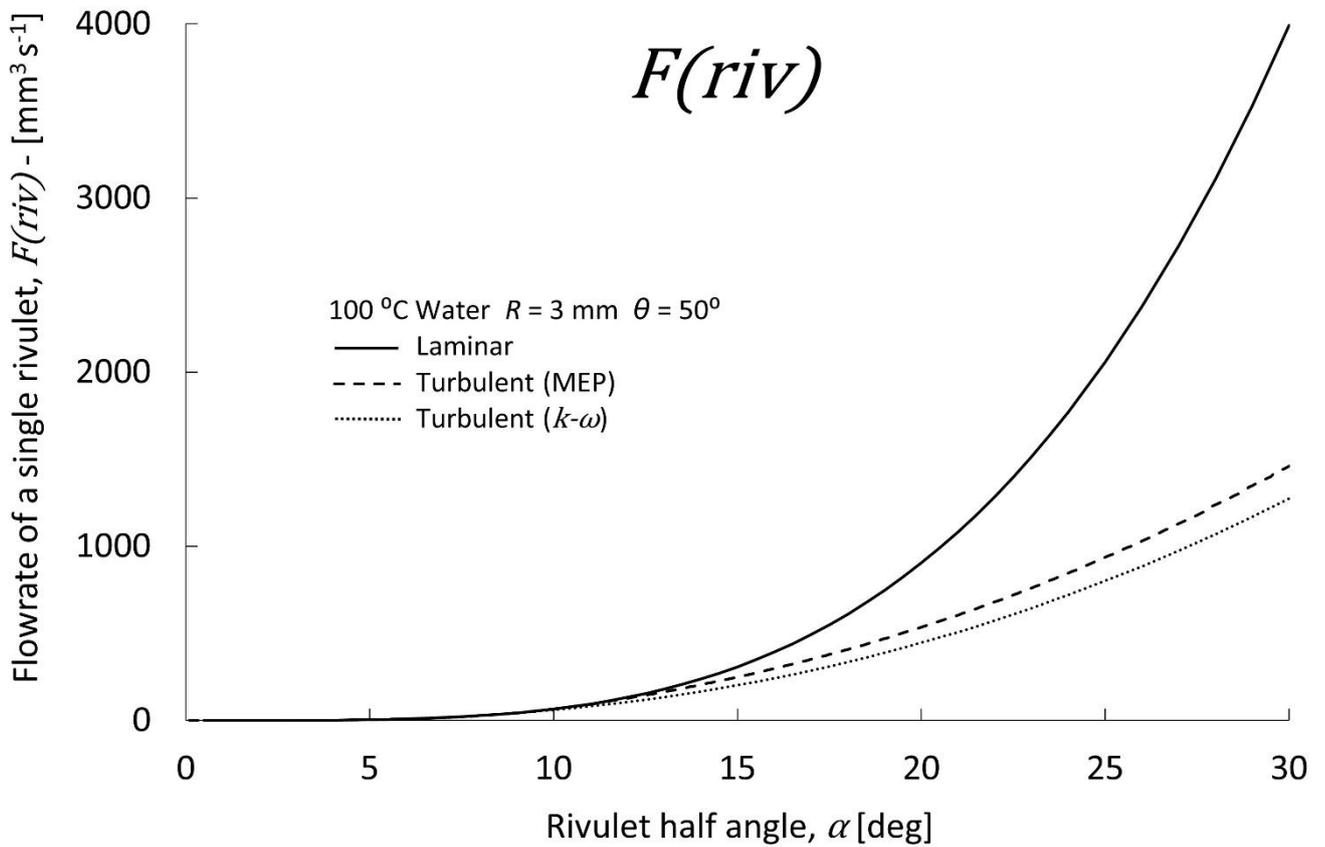

Fig. 2d   Rivulet flowrate as a function of the rivulet half angle

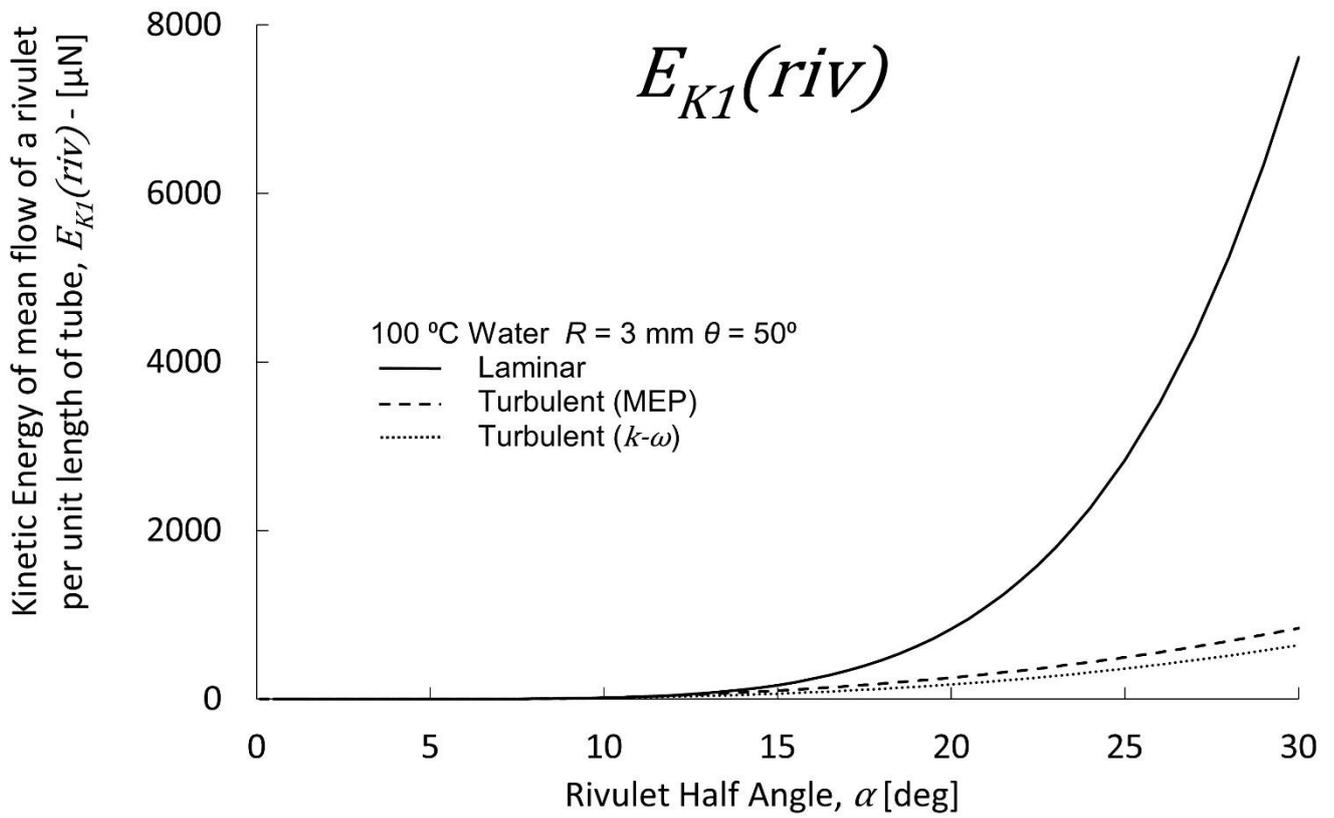

Fig. 2e Rivulet kinetic energy of the mean flow as a function of the rivulet half angle

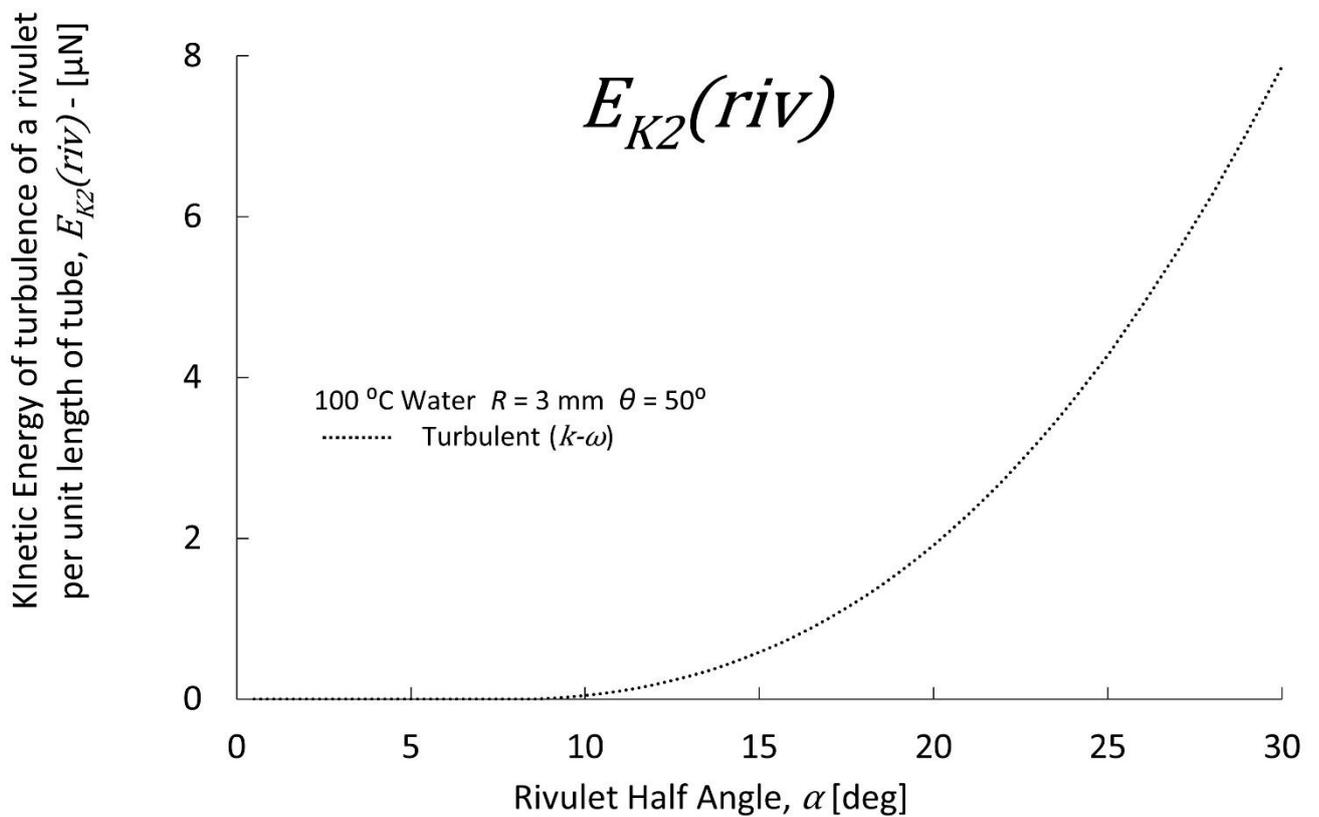

Fig. 2f Rivulet kinetic energy of turbulence as a function of the rivulet half angle

Fig. 2 Flowrates and energies of films and rivulets

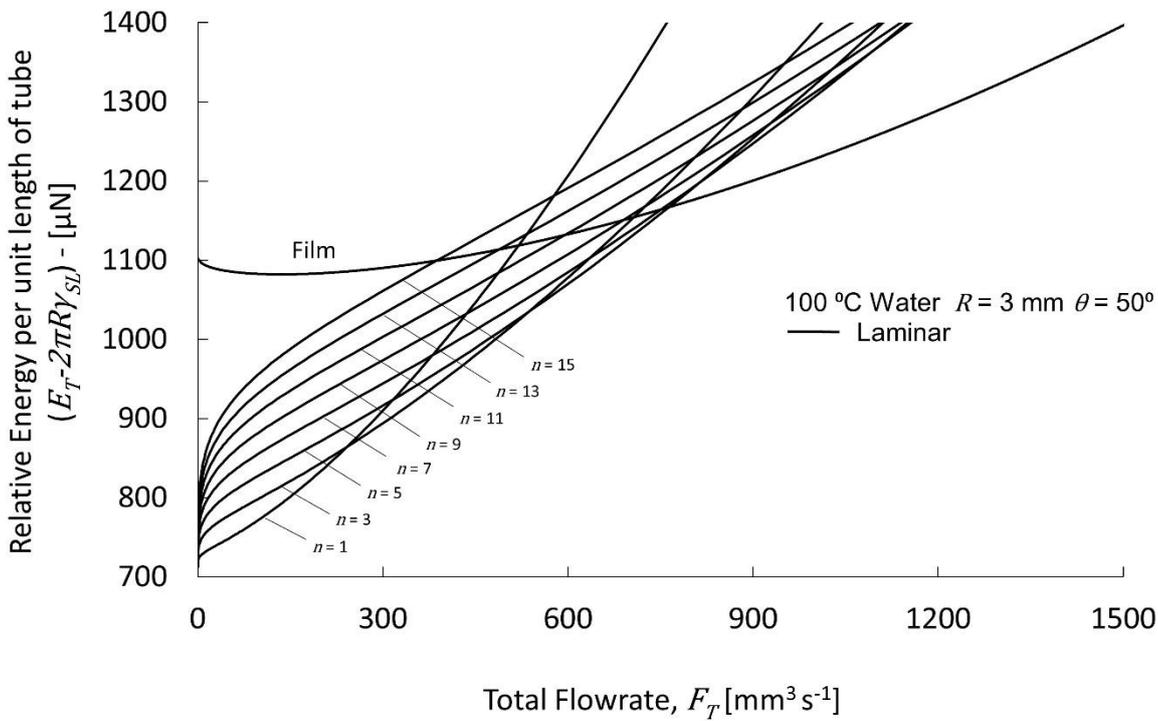

Fig. 3a  Relative energy versus flowrate for laminar flow

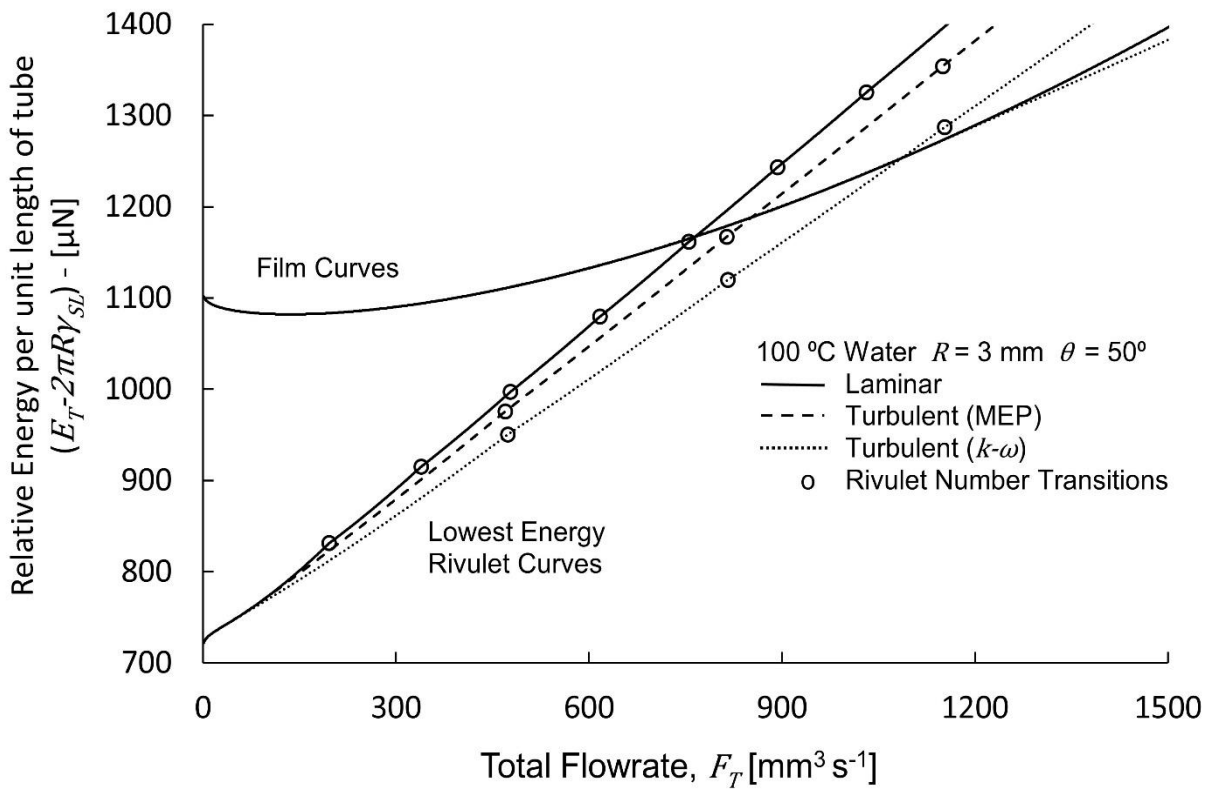

Fig. 3b  Relative energy versus flowrate for laminar and turbulent flows

Note: The Laminar and Turbulent (MEP) film curves are coincident over the flowrate range shown in this graph.

Fig. 3 Graphs of Relative energy versus flowrate

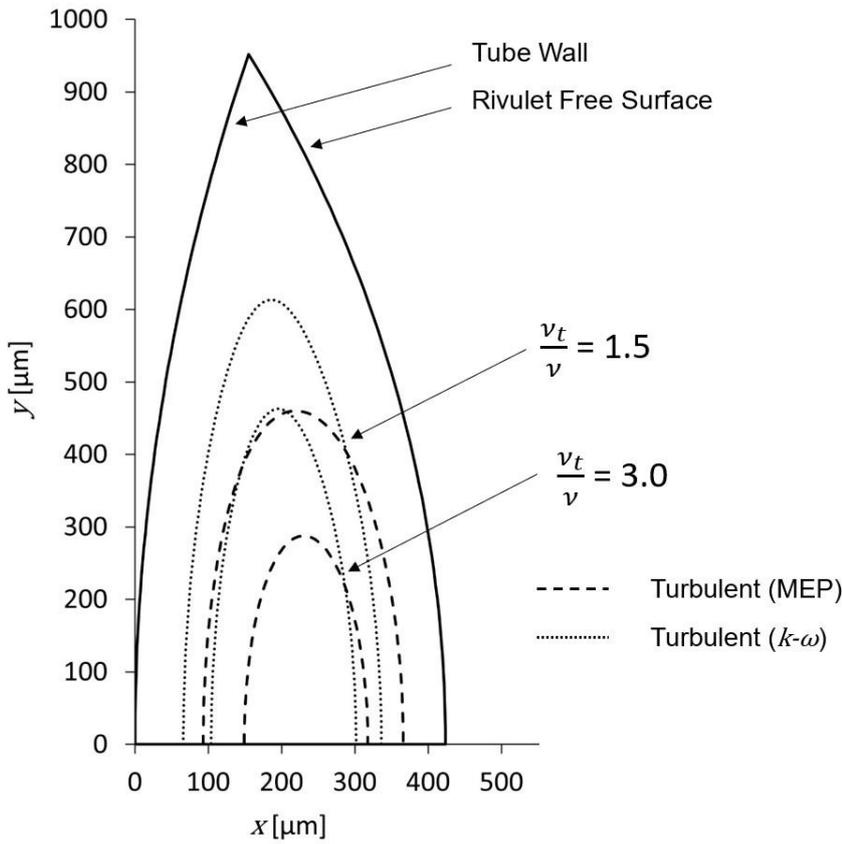

Fig. 4  Rivulet viscosity contours ($\alpha$ = 18.5º  $\theta$ = 50º  $R$ = 3 mm 100 ºC water)

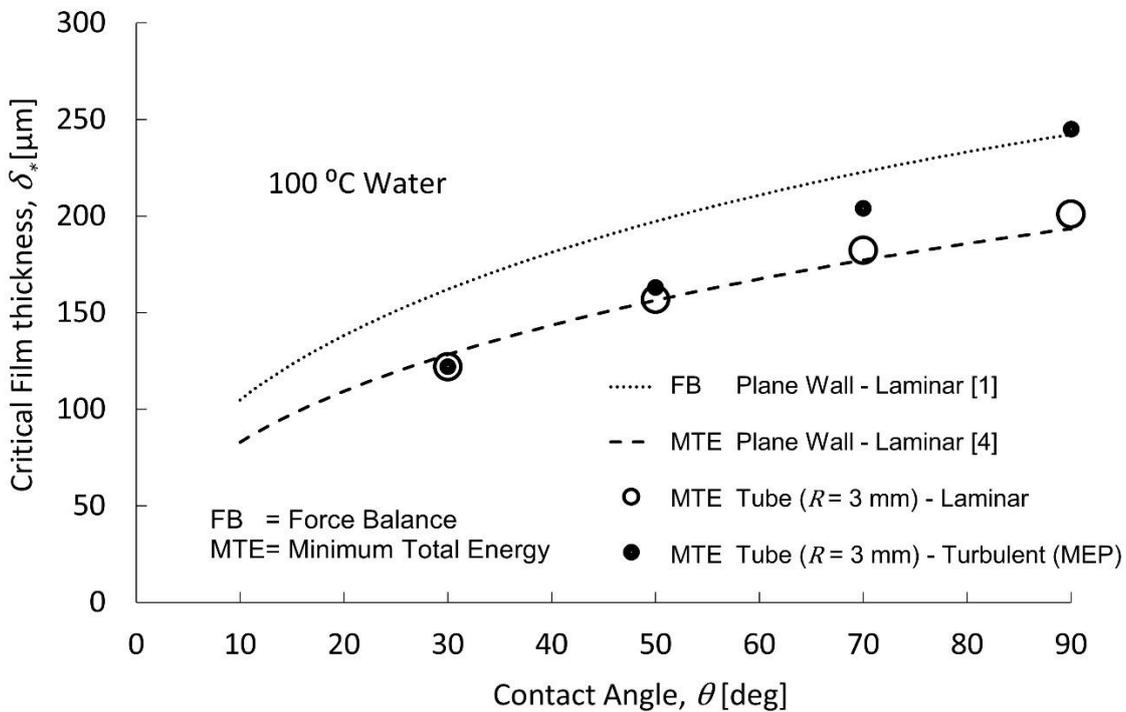

Fig. 5 Critical film thickness as a function of contact angle

Note: The equations used to calculate the laminar plane wall curves are shown in Table 1.  The plane wall FB Turbulent(MEP) curve is very close to the laminar FB curve shown in the Figure and so it has not been included on the graph (calculation presented in the Appendix)

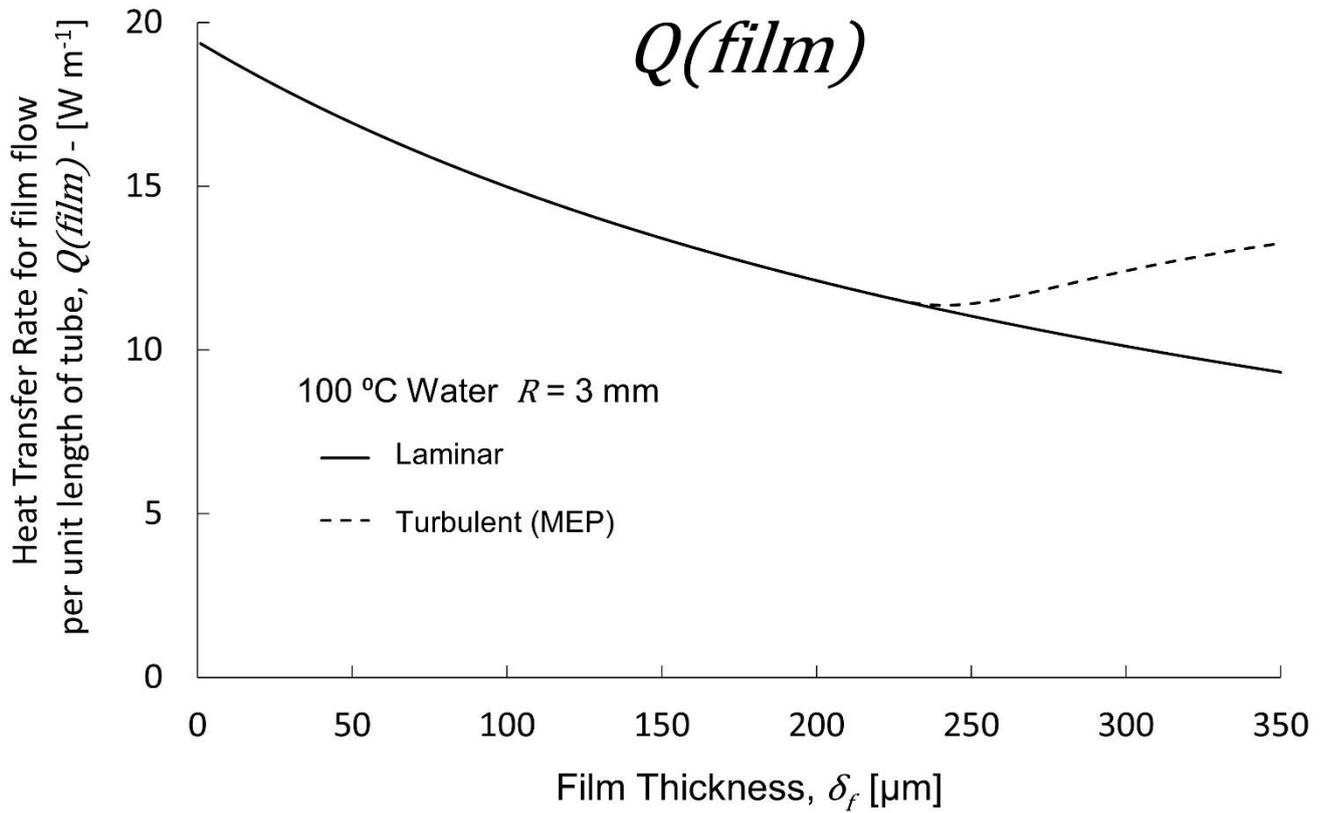

Fig. 6a Rate of heat transfer for films

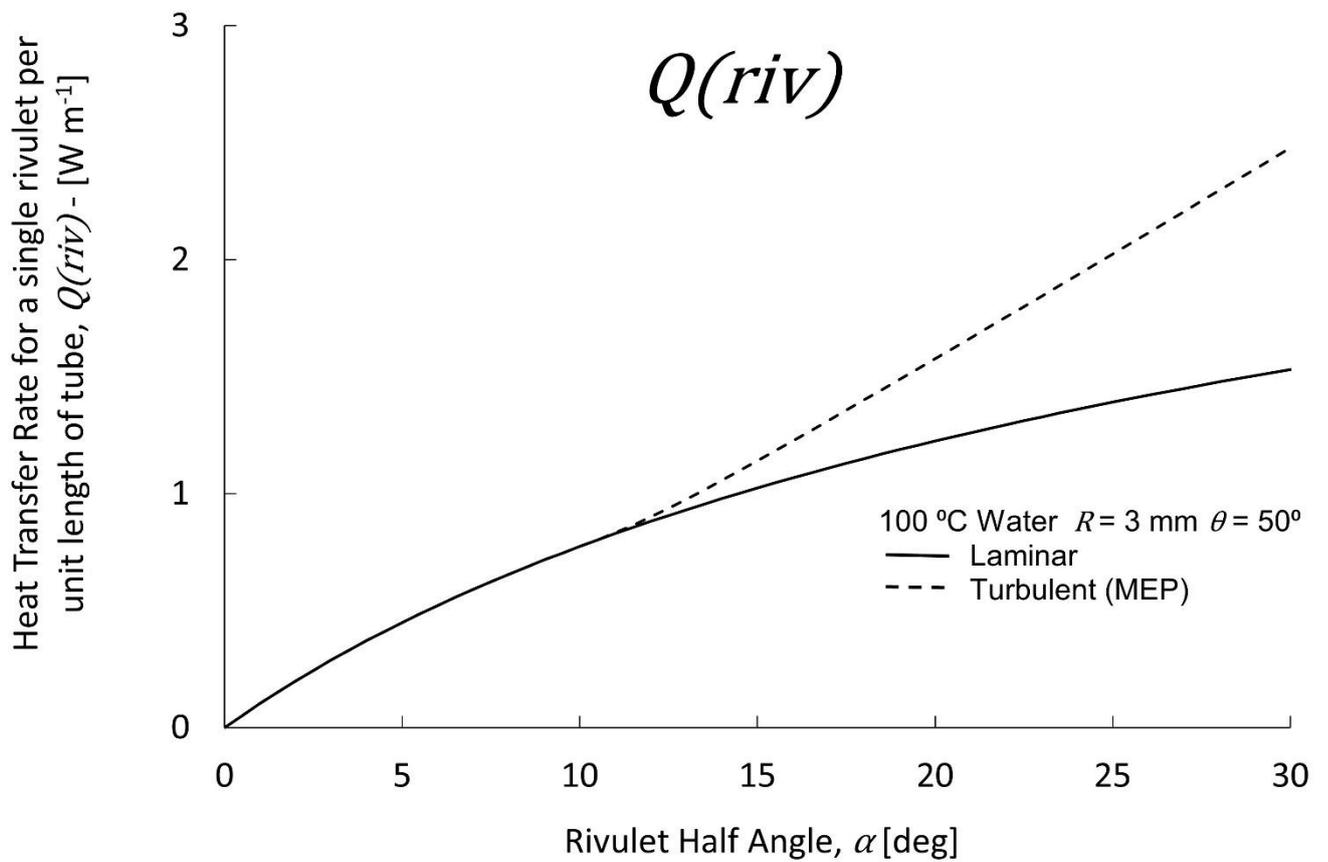

Fig. 6b Rate of heat transfer for rivulets

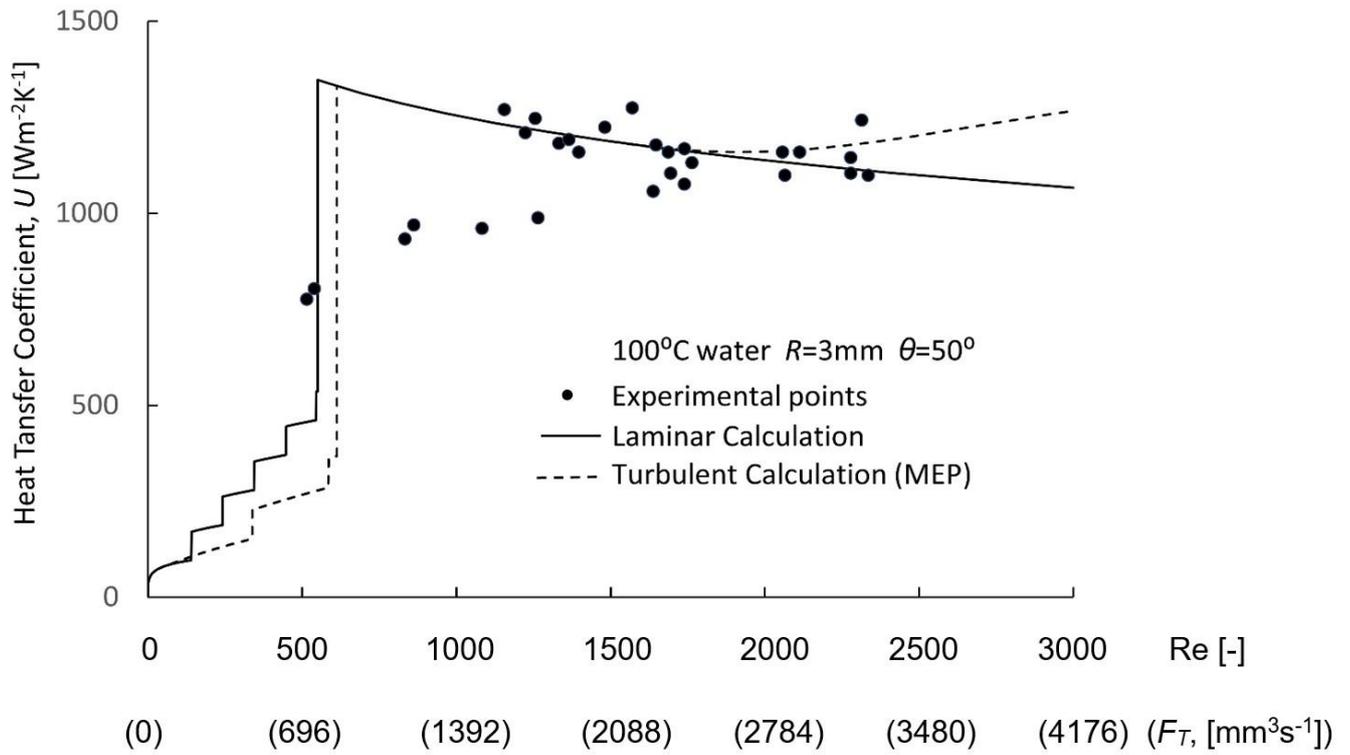

Fig. 6c  Theoretical and experimental heat transfer coefficients as a function of Reynolds Number

Fig. 6  Heat Transfer Calculations

TABLES (2 Tables)

| Equation | Reference & Notes |
|---|---|
| $\Delta_* = (1 - cos\theta)^{1/5}$ | Hartley and Murgatroyd [1] <br><br> • Plane Wall model <br> • A force balance (FB) is made at the upstream stagnation point of a dry patch. <br> • The equation shown is for laminar flow, but the force balance model can also be used with turbulent flows. |
| $\Delta_* = \left(\dfrac{3}{2}\right)^{1/5} (1 - cos\theta)^{1/5}$ | Hobler [2] <br><br> • Plane Wall model <br> • The model uses the minimum total energy (MTE) criterion <br> • Laminar Flow <br> • The model does not give any information on the geometry or spacing of the rivulets resulting from the film break-up |
| $\Delta_*^5 + (1 - cos\theta) - G(\theta)\Delta_*^3 = 0$ | Mikielewicz and Moszynski [4] <br><br> • Plane Wall model <br> • The model uses the minimum total energy (MTE) criterion <br> • Laminar Flow <br> • The radius of curvature and spacing of the rivulets resulting from film break-up can be calculated |

where

$$G(\theta) = \left(\frac{2}{3}\right)^{3/5}\left(\frac{5}{2}\right)\left(\frac{\sin\theta}{f(\theta)}\right)\left(\frac{\psi(\theta)}{\sin\theta}\right)^{3/5}\left(\frac{\theta}{\sin\theta} - \cos\theta\right)^{2/5}$$

$$f(\theta) = -\frac{1}{4}\cos^3\theta \sin\theta - \frac{13}{8}\cos\theta\sin\theta - \frac{3}{2}\theta\sin^2\theta + \frac{15}{8}\theta$$

$$\psi(\theta) = \theta\left(\frac{5}{16} + \frac{15}{4}\cos^2\theta + \frac{5}{2}\cos^4\theta\right) - \sin\theta\left(\frac{113}{48}\cos\theta + \frac{97}{24}\cos^3\theta + \frac{1}{6}\cos^5\theta\right)$$

The critical film thickness $\delta_*$ is related to the dimensionless critical film thickness $\Delta_*$ by the equation $\delta_* = \Delta_*/\Lambda$ where $\Lambda = (\rho^3 g^2/15\mu^2\gamma_{LV})^{1/5}$.
Using the physical property data presented in the Appendix, $\Lambda$ (for 100 ºC water) = 4127 m$^{-1}$.
For laminar flow on a plane wall, the minimum wetting rate $\Gamma_*$ is related to $\delta_*$ by the Nusselt equation $\Gamma_* = \rho^2 g \delta_*^3/3\mu$

Table 1 Models to calculate minimum film thickness

| θ | MTE Laminar Calculation | | | | | | MTE Turbulent Calculation (MEP) | | | | | |
|---|---|---|---|---|---|---|---|---|---|---|---|---|
| | $\delta_*$ [μm] | $F_*$ [mm$^3$ s$^{-1}$] | $n_*$ [-] | $\alpha_*$ [deg] | $\delta_{max_*}$ [μm] | $X_*$ [-] | $\delta_*$ [μm] | $F_*$ [mm$^3$ s$^{-1}$] | $n_*$ [-] | $\alpha_*$ [deg] | $\delta_{max_*}$ [μm] | $X_*$ [-] |
| 30 | 122 | 360 | 4 | 16.4 | 223 | 0.36 | 122 | 360 | 4 | 16.4 | 223 | 0.36 |
| 50 | 157 | 765 | 6 | 11.9 | 278 | 0.40 | 163 | 853 | 3 | 15.7 | 363 | 0.26 |
| 70 | 182 | 1188 | 7 | 9.4 | 326 | 0.37 | 204 | 1655 | 3 | 15.5 | 523 | 0.26 |
| 90 | 201 | 1584 | 9 | 7.1 | 350 | 0.36 | 245 | 2773 | 4 | 13.3 | 628 | 0.30 |

MTE= Minimum Total Energy

Table 2 Critical conditions for laminar and turbulent calculations ($R$ = 3 mm, 100 ºC water)